\documentclass{jfm_cup}
\usepackage{graphicx}
\usepackage{epstopdf, epsfig}
\usepackage{hyperref}
\usepackage[usenames,dvipsnames,svgnames,table]{xcolor}
\usepackage{libertine}
\usepackage{mathptmx} % comment out on mac
\usepackage[utf8]{inputenc}
\usepackage{newtxmath}
\usepackage{pgfplots}
\usepackage{natbib}

\hypersetup{
colorlinks=true,
linkcolor=blue,
urlcolor=blue,
filecolor=blue,
citecolor=blue
}

\makeatletter

\patchcmd{\NAT@citex}
  {\@citea\NAT@hyper@{%
     \NAT@nmfmt{\NAT@nm}%
     \hyper@natlinkbreak{\NAT@aysep\NAT@spacechar}{\@citeb\@extra@b@citeb}%
     \NAT@date}}
  {\@citea\NAT@nmfmt{\NAT@nm}%
   \NAT@aysep\NAT@spacechar\NAT@hyper@{\NAT@date}}{}{}

\patchcmd{\NAT@citex}
  {\@citea\NAT@hyper@{%
     \NAT@nmfmt{\NAT@nm}%
     \hyper@natlinkbreak{\NAT@spacechar\NAT@@open\if*#1*\else#1\NAT@spacechar\fi}%
       {\@citeb\@extra@b@citeb}%
     \NAT@date}}
  {\@citea\NAT@nmfmt{\NAT@nm}%
   \NAT@spacechar\NAT@@open\if*#1*\else#1\NAT@spacechar\fi\NAT@hyper@{\NAT@date}}
  {}{}

\makeatother

\makeatletter
\edef\@figdir{_pgfcache}

\catcode`\#=11
\catcode`\|=6
\newcommand{\@basicpgfpreamble}[1]{%
    \unexpanded{%
        \documentclass{standalone}^^J
        \usepackage{pgf}^^J
        \let\oldpgfimage\pgfimage^^J
        \renewcommand{\pgfimage}[2][]{\oldpgfimage[#1]{|1/#2}}^^J
    }%
}
\catcode`\#=6
\catcode`\|=12

\let\@pgfpreamble\@basicpgfpreamble

\newcommand{\setpgfpreamble}[1]{%
    \renewcommand{\@pgfpreamble}[1]{\@basicpgfpreamble{##1}\unexpanded{#1}}
}

\newcounter{@pgfcounter}
\newwrite\@pgfout
\newread\@pgfin

\newcommand{\importpgf}[3][]{%
    \IfFileExists{#2/#3}{}{\errmessage{importpgf: File #2/#3 not found}}%
    \edef\@figfile{\jobname-\the@pgfcounter}%
    \providecommand{\@writetempfile}{}%
    \renewcommand{\@writetempfile}[1]%
    {%
        \immediate\openout\@pgfout=##1%
        \immediate\write\@pgfout{\@pgfpreamble{#2}}%
        \immediate\write\@pgfout{\string\begin{document}}%
        \immediate\openin\@pgfin=#2/#3%
        \begingroup\endlinechar=-1%
            \loop\unless\ifeof\@pgfin%
                \readline\@pgfin to \@fileline%
                \ifx\@fileline\@empty\else%
                    \immediate\write\@pgfout{\@fileline}%
                \fi%
            \repeat%
        \endgroup%
        \immediate\closein\@pgfin%
        \immediate\write\@pgfout{\string\end{document}}%
        \immediate\closeout\@pgfout%
    }%
    \def\@compile%
    {%
        \immediate\write18{pdflatex -interaction=batchmode -output-directory="\@figdir" \@figdir/\@figfile.tex}%
    }%
    \IfFileExists{\@figdir/\@figfile.pdf}%
    {%
        \@writetempfile{\@figdir/tmp.tex}%
        \edef\@hashold{\pdfmdfivesum file {\@figdir/\@figfile.tex}}%
        \edef\@hashnew{\pdfmdfivesum file {\@figdir/tmp.tex}}%
        \ifnum\pdfstrcmp{\@hashold}{\@hashnew}=0%
            \relax%
        \else%
            \@writetempfile{\@figdir/\@figfile.tex}%
            \@compile%
        \fi%
    }%
    {%
        \@writetempfile{\@figdir/\@figfile.tex}%
        \@compile%
    }%
    \IfFileExists{\@figdir/\@figfile.pdf}%
    {\includegraphics[#1]{\@figdir/\@figfile.pdf}}%
    {\errmessage{Error during compilation of figure #2/#3}}%
    \stepcounter{@pgfcounter}%
}
\makeatother

\setpgfpreamble{%
\usepackage{libertine}
\usepackage{mathptmx} % comment out on mac
\usepackage[utf8]{inputenc}
\usepackage{newtxmath}
}

\newcommand{\cmd}[1]{\textbackslash\texttt{#1}}
\defcitealias{companion}{GMS}
\newcommand{\eps}{\varepsilon}

\newcommand{\JL}[1]{{\color{red}[#1  -Jake]}}
\newcommand{\RICH}[1]{{\color{cyan}[#1  -Rich]}}
\newcommand{\TOM}[1]{{\color{magenta}[#1  -Tom]}}
\newcommand{\vect}[1]{\boldsymbol{#1}}
\newcommand{\grad}{\bnabla}
\newcommand{\cc}{\mathrm{c.c.}}
\newcommand\Ri{\mbox{\textit{Ri}}} % Richardson number
\newcommand\Ra{\mbox{\textit{Ra}}} % Rayleigh number
\newcommand\shiftrefl{\mathcal{S}} % shift and reflect symmetry
\newcommand\zrot{\varOmega} % z-axis rotation symmetry
\newcommand\spanrefl{\mathcal{Z}} % spanwise reflect symmetry

\title{Stably stratified exact coherent structures in shear flow: the effect of Prandtl number}
\shorttitle{Effect of Prandtl number on stably stratified exact coherent structures}

\shortauthor{J. Langham, T.  S. Eaves and R. R. Kerswell}
\author{Jake Langham\aff{1}
  \corresp{\email{j.langham@bristol.ac.uk}},
  Tom S. Eaves\aff{2}
 \and Rich R. Kerswell\aff{1,3}}

\affiliation{\aff{1}School of Mathematics, University of Bristol,
Bristol, BS8 1TW, UK
\aff{2}Department of Mathematics, University of British Columbia, 1984
Mathematics Rd., Vancouver, BC, V6T 1Z2, Canada
\aff{3}Centre for Mathematical Sciences, University of Cambridge, Cambridge,
CB3 0WA, UK}

\begin{document}

\maketitle

\begin{abstract}
We examine how known unstable equilibria of the Navier-Stokes equations in plane
Couette flow adapt to the presence of an imposed stable density difference
between the two boundaries for varying values of the Prandtl number $\Pran$,
the ratio of viscosity to density diffusivity, and fixed moderate Reynolds
number, $\Rey=400$.  In the two asymptotic limits $\Pran \to 0$ and $\Pran \to
\infty$, it is found that such solutions exist at arbitrarily high bulk
stratification but for different physical reasons. In the $\Pran \to 0$ limit,
density variations away from a constant stable density gradient become
vanishingly small as diffusion of density dominates over advection, allowing
equilibria to exist for bulk Richardson number $\Ri_b \lesssim
O(\Rey^{-2}\Pran^{-1})$. % \gg 1$. 
%
%Alternatively, in the $\Pran \to \infty$ limit, the flow becomes
Alternatively, at high Prandtl numbers, density becomes
homogenised in the interior by the dominant advection which creates strongly
stable stratified boundary layers 
%of $O(\Pran^{-1/3})$ thickness 
that recede into the wall as $\Pran\to\infty$.
%In this scenario, equilibria can exist for $\Ri_b \lesssim O(\Pran^{5/9})$ at
%fixed $\Rey$, as the stratification and the flow essentially separate.
%
In this scenario, the density stratification and the flow essentially decouple,
thereby mitigating the effect of increasing $\Ri_b$.
An asymptotic analysis  is presented in the passive scalar regime $\Ri_b
\lesssim O(\Rey^{-2})$, which reveals $O(\Pran^{-1/3})$-thick stratified
boundary layers with $O(\Pran^{-2/9})$-wide eruptions, giving rise to
density fingers of $O(\Pran^{-1/9})$ length and $O(\Pran^{-4/9})$ width
that invade an otherwise homogeneous interior.
Finally, increasing $\Rey$ to $10^5$ in this
regime reveals that {\em interior} stably stratified density layers can form
away from the boundaries, separating well-mixed regions. 
%\RICH{The connection of these
%to step formation and layering seen in turbulent stratified flows is discussed. - remove? not sure we do this now}
%
\end{abstract}

\section{Introduction}
The identification of numerous unstable invariant solutions in canonical shear
flows has played a key role in the modern understanding of transitionally
turbulent fluids.  These states, most commonly known as \emph{exact coherent
structures (ECS)} underpin a deterministic picture of turbulence in which the
flow is viewed as a complicated trajectory through a state space of velocity
fields that satisfy the Navier-Stokes
equations~\citep{Kerswell2005,Eckhardt2007,Kawahara2012}.  The ECS are simple
invariant solutions, equilibria, travelling waves and (relative) periodic
orbits, that organise this state space through their stable and unstable
manifolds.
%These states, most commonly known as \emph{exact coherent
%structures (ECS)} sit apart from laminar flow in the phase space of the
%Navier-Stokes dynamical system, either embedded in the turbulent attractor or
%along the laminar-turbulent boundary. 
Though their unstable eigendirections prohibit them from being physically
realised, turbulent flows may pass nearby to them. Consequently, the dynamical
influence of ECS on the flow via their invariant manifolds is experimentally
observable~\citep{Suri2017,Suri2018} and may be investigated in detail by
high-resolution direct numerical simulations
(DNS)~\citep{Kerswell2007,Gibson2008,Cvitanovic2010,Budanur2017}. 
As such, they have proven useful in investigating both the route to
turbulence~\citep{Itano2001,Skufca2006,Schneider2007,Kerswell2007,Viswanath2009,Mellibovsky2009,Schneider2010,Duguet2010,Pringle2012}
and its subsequent
features~\citep{Gibson2008,Willis2013,Chandler2013,Lucas2015,Budanur2017}.
%

%The majority of ECS discovered in shear flows are realisations of the
%\emph{self-sustaining process (SSP)} mechanism of \citet{Waleffe1997} or
%(equivalently) finite Reynolds number realisations of the asymptotic
%\emph{vortex-wave interaction (VWI)} theory of \citet{Hall1991} 
%\citep[see][for an exception]{Smith1982,Deguchi2013}. Such solutions consist of a nearly
%streamwise independent roll-streak/vortex structure sustained by a
%streamwise-modulating wave: the wave prevents the rolls from decaying, and the
%rolls produce the streaks, which in turn suffer a marginal instability which
%results in a wave. 

Past work has focussed largely on understanding ECS in simple shear flows in
channels and pipes whose physics can be parametrised by a single dimensionless
variable: the Reynolds number $\Rey$.
However, it is common in industrial, environmental and astrophysical flows, for
the fluid to be subject to additional (buoyancy) forces due to gradients in
density or temperature. In order to understand the role of ECS in such flows,
one must therefore catalogue the effect of introducing this extra physics onto
these states in homogeneous shear flows.
Stratified flows require that at least two additional parameters be accounted
for; the bulk Richardson number $\Ri_b$ controls the global level of
stratification and the Prandtl number $\Pran$ 
sets the ratio of viscosity to density diffusion.
%dictates the balance
%between advective versus diffusive density transport \RICH{that's the definition of the Peclet number = RePr - Pr is the ratio of viscous to density diffusion}
%
While identifying ECS within this augmented parameter space presents a sizeable
challenge, it is also an opportunity, since many physically relevant regimes
remain unexplored.

A number of prior studies have investigated ECS in the stably
stratified setting , where the bulk density gradient acts to suppress
energetically unfavourable vertical motions.  \citet{Eaves2015} considered the
effect of weak stratification on a particular state lying on the laminar-turbulent
boundary in stratified plane Couette flow, providing a scaling argument which
demonstrated that the usual physical structure of ECS 
\citep[the SSP/VWI mechanism, see \S\ref{sec:ecstheory}
and][]{Waleffe1997,Hall1991,Hall2010} is disrupted when $\Ri_b =
O(\Rey^{-2})$.
This was subsequently confirmed asymptotically by both \citet{Deguchi2017} and
\citet{Olvera2017}. The latter study conducted a substantial survey of the
$\Rey$-$\Ri_b$ space for plane Couette flows and $\Pran=1$, 
%finding that the effect of stratification is to generate a connection between
%the upper and lower branch solutions known in the unstratified system.
%
continuing solution branches for states at $\Ri_b>0$ that connect known
unstratified ECS pairs, which can then be tracked into $\Ri_b < 0$ (unstable
stratification), to bifurcations of sheared Rayleigh-B\'{e}nard solutions. They
noted that a key effect of increasing stratification is to encourage ECS to
localise in one or more directions, conjecturing the existence of
fully-localised states when $\Ri_b = O(\Rey^0)$.
Finally, Lucas \emph{et al.} discovered ECS 
%\citep{Lucas2017b} and
%periodic orbits \citep{Lucas2017} 
in two-dimensional stratified shear flow with a sinusoidal body forcing
(Kolmogorov flow).  They observed that both equilibria and turbulent DNS form
shear and density layers originating from a sequence of linear `zig-zag'
instabilities in the (sinusoidal) base flow \citep{Lucas2017b}, and converged
periodic orbits to study how stratification affects mixing efficiency
\citep{Lucas2017}.

Despite this recent progress in the understanding of stratified ECS, a
connection with realistic fully turbulent stratified shear flow remains
unresolved (though \cite{Lucas2018} arguably see traces of underlying ECS in
recent stratified plane Couette DNS results).
In part this is due to the Prandtl number; prior stratified ECS
studies have only dealt with unit Prandtl numbers
\citep{Eaves2015,Deguchi2017,Olvera2017,Lucas2017b,Lucas2017} and full DNS
of stratified shear flow turbulence is restricted in the Prandtl numbers it can
access by issues of numerical resolution. However, Prandtl numbers observed in
nature span a substantial range of scales, from as low as $10^{-7}$ (inside
stars) to up to at least $10^{23}$ (Earth's mantle).  On a more terrestrial
level, temperature in air exhibits a Prandtl number around 0.7 (which is
certainly attainable via DNS), whereas salt in water has $\Pran \approx 700$
(currently unattainable via DNS, save in the $\Ri_b\to 0$ limit).

This latter case of saltwater shear flow turbulence is of particular interest
since the structure of turbulence in the oceans governs the mass transport
across its depth \citep{Munk1966}. As such, significant effort has been put into
explaining the mass transport via modelling of the mixing of the density field
that the turbulence causes \citep{Linden1979,Caulfield2003,Sutherland2019}.
Additionally, it has been observed that stratified shear flow turbulence
typically organises into well-mixed constant density layers of large vertical
extent separated by relatively thin interfacial regions between each layer
\citep{Turner1973,Gregg1980}. 
%
% [not all (or even any) lab experiments necessarily relevant since Phillips mechanism doesn't seem to generate layers]
%This phenomenon is confirmed in the laboratory and by
%numerical simulations: see e.g.\ the review of \cite{Thorpe2016}.
%
Layers may appear spontaneously at turbulent Reynolds numbers and
%are not linked 
have been reproduced in a number of singly-diffusive experimental flows
\citep{Ruddick1989,Park1994,Holford1999,Oglethorpe2013,Thorpe2016}. (The doubly-diffusive case is also
important, but beyond our scope.)
%any double-diffusive process \citep{Holford1999,Oglethorpe2013}. 
The vertical extent of the well-mixed regions is linked to the flow speed and
the buoyancy frequency \citep{Thorpe2016}, but the detailed structure of the
layers and the interfaces between them remains an open question, although the
`sharpness' of the interfaces appears to increase as the Prandtl number
increases \citep{Zhou2017b}.
%Layering can be explained via an anti-diffusive mass flux structure 
%that causes sufficiently strong local density gradients to become unstable \citep{Phillips1972,Posmentier1977,Balmforth1998}.
%%at strong enough stratification strength \citep{Phillips1972,Balmforth1998}.  
 
Drawing motivation from the above examples and the recent interest in stratified
ECS, our plan here is to complement the stratified plane Couette flow study of
\citet{Olvera2017}, by probing the $\Ri_b$-$\Pran$ space at fixed Reynolds
number. We are particularly interested here in the as-yet unexplored limits
$\Pran \ll 1$ and $\Pran \gg 1$, which are relevant for understanding
astrophysical and geophysical flows respectively.
The key questions are: (\emph{a}) how strong can the bulk stratification be for
ECS still to exist; and (\emph{b}) what is the effect of $\Pran$ on the
structure of solutions? The outline of the paper is as follows.  In
\S\ref{sec:setting}, we detail the physical system, the numerical methods used
and the necessary underlying theory (SSP/VWI) for exact coherent states in shear
flows.
We then take the principal solution of \cite{Olvera2017} and perform parameter
continuation in $\Ri_b$ for a wide range of fixed Prandtl numbers
(\S\ref{sec:results}), allowing us to identify the range of global
stratification that states exist for, before splitting our analysis into the
low-$\Pran$ (\S\ref{sec:lowpr}) and high-$\Pran$ (\S\ref{sec:highpr}) limits.
In the low-$\Pran$ case, we identify and study an asymptotic solution branch
onto which all states collapse.
For high-$\Pran$ states, we intriguingly observe the density field splitting
into a homogenised region with highly stratified boundary layers at the walls,
punctuated by jets of advected density.
We conduct an analysis in \S\ref{sec:ri0} of its structure in the limit of weak
stratification, before proceeding in \S\ref{sec:increasingri} to consider
increasing $\Ri_b$. As states leave the weakly stratified regime, their velocity
perturbations retreat from the walls, apparently inhibited by the presence of
high stratification.
Section \ref{sec:discussion} concludes with a final discussion of these results
and presents some preliminary observations of interfaces developing in the
channel interior as $\Pran$ increases at high Reynolds number.

\section{Setting}\label{sec:setting}
\subsection{Stratified plane Couette flow}
\label{sec:spcf}
The setting for this study  is  plane Couette flow, which is the flow of an
incompressible viscous fluid between two infinite parallel plates at $y=\pm h$,
moving with velocities  $\pm U \vect{e}_x$.  The flow velocity is represented as
$\vect{u}:=u \vect{e}_x +v \vect{e}_y +w \vect{e}_z$ (where $\vect{e}_x$,
$\vect{e}_y$ and $\vect{e}_z$ indicate the \emph{streamwise}, \emph{wall-normal}
and \emph{spanwise}
directions respectively) and obeys no-slip boundary conditions at the plates.
Gravity is taken to act in the wall-normal direction, $\vect{g} = -g
\vect{e}_y$. Fixed, but differing densities $\rho_0 \mp \Delta_\rho$ are imposed
at the plates $y=\pm h$, where $\Delta_\rho \ll \rho_0$ so the Boussinesq
approximation can be used. The governing equations for the fluid velocity
$\vect{u}$, pressure $p$ and density $\rho$ are 
\begin{subequations}
\begin{align}
    \frac{\partial\vect{u}}{\partial t} + \vect{u}\bcdot\grad\vect{u} & =
    -\grad p + \frac{1}{\Rey}\nabla^2 \vect{u} - \Ri_b \rho\,
	\vect{e}_y,\label{eq:momfull}\\
	\grad \bcdot \vect{u} & = 0,\label{eq:incompfull}\\
    \frac{\partial\rho}{\partial t} + \vect{u} \bcdot \grad \rho & 
	= \frac{1}{\Rey\Pran} \nabla^2 \rho,\label{eq:stratfull}%
\end{align}%
\label{eq:boussinesq}%
\end{subequations}%
where $U$, $h$ and $\Delta_{\rho}$,  have been used to non-dimensionalise the
system (so that the total dimensional density is $\rho_0+\Delta_{\rho}\rho$, for
example). The dimensionless parameters $\Rey$, $\Pran$ and $\Ri_b$ are the
\emph{Reynolds}, \emph{Prandtl} and \emph{bulk Richardson} numbers respectively,
defined as
\begin{equation}
	\Rey \coloneqq \frac{Uh}{\nu},\quad \Pran \coloneqq \frac{\nu}{\kappa},\quad
	\Ri_b \coloneqq
	\frac{\Delta_\rho}{\rho_0}\frac{g h}{U^2},
\end{equation}
where $\nu$ is the kinematic viscosity of the fluid and $\kappa$ is the
diffusivity of some density-affecting agent, such as temperature or salt
content. The accompanying boundary conditions are $u = \pm 1$, $v = 0$, $w = 0$,
$\rho = \mp 1$ at the walls $y = \pm 1$ which, along with
Eqs.~\eqref{eq:momfull}--\eqref{eq:stratfull},  admit a steady base flow with
linear velocity and density profiles:
\begin{equation}
\vect{u} = y\vect{e}_x, \quad p = \Ri_b y^2 / 2, \quad \rho = -y.
\end{equation}

\subsection{SSP/VWI mechanism}
\label{sec:ecstheory}%
In this study, we analyse unstable equilibrium solutions to
Eqs.~\eqref{eq:momfull}--\eqref{eq:stratfull} that differ from the basic
laminar flow.  Almost all such states in homogeneous shear flows where the base state is
linearly stable arise physically due to the  \emph{self-sustaining process
(SSP)} mechanism proposed by~\cite{Waleffe1997}, which was subsequently
recognised to be the finite-$Re$ manifestation of the
\emph{vortex-wave-interaction (VWI)} theory \citep{Hall1991,Hall2010}. (The
state proposed by \cite{Smith1982} and recently found numerically by
\cite{Deguchi2013} is one exception.) 

The SSP/VWI process relies on three flow structures, \emph{rolls, streaks}
and \emph{waves}, sustaining each other against dissipation in a closed loop.
Streamwise rolls advect the underlying shear causing the streamwise velocity to
vary in the spanwise direction, thereby creating \emph{streaks}, sustained
patches of negative or positive streamwise velocity adjustments to the basic
shear.  If these streaks have large enough amplitude, they are unstable to
streamwise-varying waves, which, via the quadratic nonlinearity in
Eq.~\eqref{eq:momfull}, can then energise the original streamwise rolls to
close the loop. This process is covered in great detail
elsewhere~\citep{Hamilton1995, Waleffe1997, Hall1991, Hall2010}. At
asymptotically high Reynolds number, these states have the following structure:
\begin{subequations}
\begin{align}
u &=  \quad \;\;\;\; \bar{u}(y,z)+ \ldots + \delta^\prime(\Rey)[ U(y,z)\mathrm{e}^{ikx} + \cc]+ \ldots \label{eq:vwi1}\\
v &= \Rey^{-1} \bar{v}(y,z)+ \ldots + \delta^\prime(\Rey)[ V(y,z)\mathrm{e}^{ikx} + \cc]+ \ldots \label{eq:vwi2}\\
w &=  \Rey^{-1}\bar{w}(y,z)+ \ldots + \delta^\prime(\Rey)[ W(y,z)\mathrm{e}^{ikx} + \cc]+ \ldots \label{eq:vwi3}\\
p &= \Rey^{-2}\bar{p}(y,z)+ \ldots + \delta^\prime(\Rey)[ P(y,z)\mathrm{e}^{ikx} + \cc]+ \ldots \label{eq:vwi4}
\end{align}
\end{subequations}
where $\delta^\prime(\Rey)=\Rey^{-7/6}$, except in the $O(\Rey^{-1/3})$-thick  critical
layer (located where the base flow vanishes for equilibria), where it is
$\Rey^{-5/6}$ \citep{Hall2010}. Here,
\begin{equation}
    \bar{q} \coloneqq \frac{1}{L_x}\int_0^{L_x} q\,\mathrm{d}x
    \label{eq:xaverage}
\end{equation}
denotes the streamwise average of a given flow field $q$.
In these expressions, $(\bar{u},0,0)$ represents the streak-modified base shear,
$(0,\bar{v},\bar{w})$ the streamwise rolls and $(U,V,W)$ the amplitudes of the
streamwise-varying wave field.

Although strictly valid only in the high-$\Rey$ limit, VWI captures the physics
of SSP remarkably well, all the way down to transitionally turbulent
$\Rey$~\citep{Hall2010}, which is the regime considered in the bulk of this study.  
The theory has been extended in the stratified setting by
\cite{Hall2012}, to incorporate plane Couette flows with a buoyancy force aligned
with the streamwise direction. Later, \cite{Deguchi2017} and \cite{Olvera2017}
considered the vortex-wave interaction for the situation herein with buoyancy
in the wall-normal direction, presenting numerical calculations at varying $\Rey$
and fixed $\Pran=1$. 
%We shall complement this work by looking at the effect of varying $\Pran$ at
%fixed $\Rey$. 
In this study, we use the VWI scalings in Eqs.~\eqref{eq:vwi1}--\eqref{eq:vwi4}
to look at the effect of varying $\Pran$ on states at fixed $\Rey$.

\subsection{Numerical methods}
The numerical solutions presented below were obtained using two separate
pseudo-spectral codes that compute stratified plane Couette equilibria in the
box $(x,y,z) \in [0, \,L_x]\times[-1,1]\times[0, \,L_z]$, with periodic
streamwise and spanwise directions of lengths $L_x$ and $L_z$ respectively.
The first was a bespoke code that computes steady solutions  of
Eqs.~\eqref{eq:momfull}--\eqref{eq:stratfull}, using the standard
Newton-Raphson method and LU decomposition to invert the discretised system
Jacobian.
%(with multiple-core cpus and multithreading of linear algebra
%software, this approach can routinely handle $O(10^5)$ degrees of freedom).
%
The second was a modified version of the Channelflow 2.0 DNS software
\citep{Channelflow} that employs a matrix-free iterative solver to find
invariant solutions via two standard methods: Stokes
preconditioning~\citep{Tuckerman1989,Tuckerman2019} and time
integration~\citep{Viswanath2007,Gibson2008}.  The former method was used to
compute $\Pran \lesssim 10$ states and the latter was used for higher $\Pran$.

Provided enough computer memory is available, a direct solver is often
preferable to the iterative strategy, since a single LU decomposition per Newton
step (using highly optimised parallel linear algebra routines) is typically
cheaper than multiple time integrations, though the recent parallelisation of
the Channelflow code has made time integration more attractive and paves the
way for fast ECS computations with very high spatial
resolutions~\citep{Channelflow}. 
To minimise its memory footprint, our direct solver code restricts solutions to
a symmetry subspace, thereby reducing the size of the basis set needed to
represent the states at a given spatial resolution: see
Appendix~\ref{appendix:numerics} for details. 
Following the previous work \citep{Deguchi2017, Olvera2017}, the symmetries
imposed (in addition to $L_x$- and $L_z$-periodicity) are: `shift-and-reflect'
$\shiftrefl$, rotation about the $z$-axis $\varOmega$ and spanwise reflection
$\spanrefl$, defined by
\begin{subequations}
\begin{align}
    \shiftrefl : [u, v, w, p, \rho](x,y,z) & \mapsto [u, v, -w, p, \rho]\left(x + L_x/2, y, L_z/2 -
    z\right),  \label{eq:shiftrefl}\\
    \zrot : [u, v, w, p, \rho](x,y,z)       &\mapsto[-u, -v, w, p, -\rho]\left(-x,-y,z\right), \\
    \spanrefl :[u, v, w, p, \rho](x,y,z) &\mapsto  [u, v, -w, p,
    \rho]\left(x,y,-z\right).\label{eq:spanrefl}
\end{align}%
\end{subequations}%
Below, streamwise-averaged quantities will often be plotted which, because of 
%
% $\zrot$ reduces to a wall-normal reflection
% in the line $y=0$ {\color{red} not strictly TRUE  as $u \rightarrow -u$ RRK} and
%
$\shiftrefl$, possess an additional spanwise reflection
symmetry in the line $z=\pi/4$. This implies that the cells
$[-1,1]\times[0,L_z/2]$ and $[-1,1]\times[L_z/2,L_z]$ are identical and obey
reflection symmetries along the lines $z = \pi/4$ and $z = 3\pi/4$ respectively.

The majority of the results presented involve a single
$\shiftrefl$-, $\zrot$- and $\spanrefl$-symmetric solution family described at
the beginning of \S\ref{sec:results} and were computed with the direct solver in
a domain with $L_x = 2\pi$ and $L_z = \pi$, at $\Rey=400$. 
%The states demanding the
%highest numerical resolution were those in the high $\Pran$ range, requiring
%$O(10^5)$ degrees of freedom and up to $\approx 400$Gb of computer memory.
%
Channelflow was later employed to obtain the additional
\S\ref{sec:othersolns} equilibria in a box with $L_x = 2\pi / 1.14$, $L_z =
2\pi/2.5$, $\Rey = 400$ and to cross-check a sample of solutions from the
direct solver.
%, using up to $O(10^6)$ spectral coefficients in the $\Pran = 70$ cases. 
Typical resolutions achieved in both cases are mentioned in
Appendix~\ref{appendix:numerics}.

Both codes are coupled to a pseudo-arclength continuation
algorithm that enables straightforward computation of solution families
parametrised by $\Rey$, $\Ri_b$ or $\Pran$. 
This allows us to track solution curves which, for the ECS studied herein, 
arise via saddle-node bifurcations \citep{Gibson2009} with
distinct \emph{upper} and \emph{lower} branches. (The lower branch lies closest
to the base state in any reasonable metric e.g.\ energy, dissipation,
amplitude.)
Since multiple equilibria may exist at the same point in parameter space, we use the
`surplus' mean wall stress
\begin{equation}
	\tau_y \coloneqq \frac{1}{L_x L_z}\int_0^{L_x}\!\!\int_0^{L_z}
	\frac{\partial u}{\partial y}\bigg|_{y=1} \,\mathrm{d}x\,\mathrm{d}z - 1,
	\label{eq:wallstress}
\end{equation}
to distinguish between them (where $\tau_y=0$ for the base flow).

\section{Results}
\label{sec:results}
This study focusses mainly on a lower branch equilibrium solution originally
discovered by~\citet{Itano2009}, who computed it via homotopy from a solution
in an unstably stratified channel with stationary walls
and~\citet{Gibson2009}, who independently converged it from snapshots of a
transitionally turbulent flow simulation. We follow the nomenclature
used by~\citet{Gibson2009}, whose search identified $13$ different
equilibria and named them $EQ_1$--$EQ_{13}$.  Our chosen solution is $EQ_7$ and
its upper branch counterpart $EQ_8$. In the vertically
stratified case with $\Pran=1$, $EQ_7$ was recently studied
by~\citet{Olvera2017}, who observed that it lies on the so-called \emph{edge
manifold} separating laminar and turbulent dynamics and is therefore likely to
play a role in organising the transition process. In \S\ref{sec:othersolns} we
briefly address other equilibrium solutions.

\begin{figure}
	%\includegraphics[width=5.3in]{figures/eq7.jpg}
	%\input{eq7.pgf}
	%\importpgf{.}{eq7.pgf}
	\includegraphics{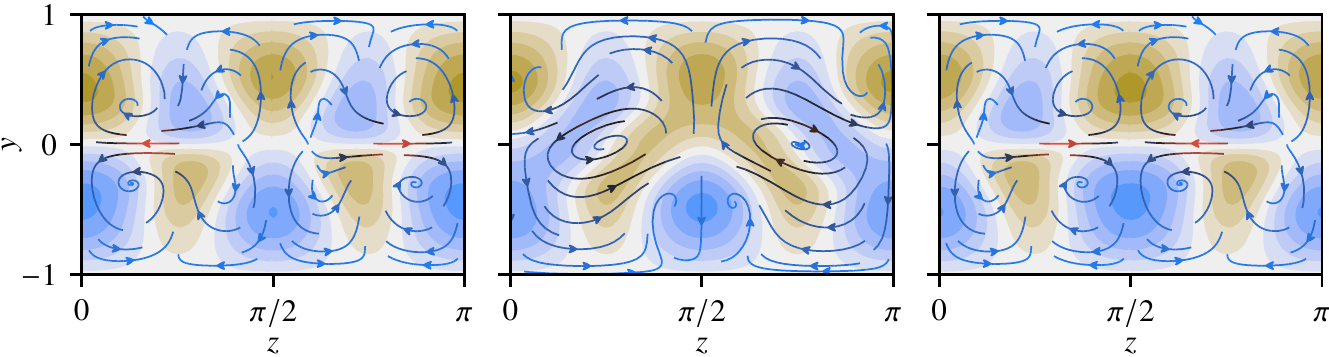}
    \caption{%
    Spanwise cross-sections of the $EQ_7$ state at $\Rey = 400$ without
    stratification.  The plots show the streamwise velocity perturbation
    $\hat{u} := u - y$, plotted using $11$ equally-spaced contour intervals
    between $-0.22$ (brown) and $0.22$ (blue). The middle interval is centred at
    zero (white).  Overlaid are streamlines of the $(v,w)$-field, coloured with
    a linear gradient from blue to red according to the magnitude of in-plane
    velocity, which lies in the interval $[0, 0.077)$.  Three $(y, z)$-slices
    are given, at $x = 0$, $x = \pi / 2$ and $x = \pi$ (left to right), covering
    half the computational domain.%
            %The remainder of the
            %solution ($\pi < x \leq 2\pi$) is dictated by the $\zrot$ symmetry
            %and $L_x$-periodicity.
		}%
	\label{fig:eq7}%
\end{figure}

Figure~\ref{fig:eq7} shows $yz$-contours of the base flow perturbation
$\hat{u} \coloneqq u - y$, for $EQ_7$
with no imposed stratification.
%
% Fig 1
%
Also shown are $(v,w)$-streamlines, coloured from blue to red according to the
speed of the flow in the plane. The leftmost panel is a slice through $x = 0$.
It contains two rows of four streaks, separated by the centre line $y = 0$. The
in-plane streamlines provide insight into the formation of the streaks.  For
example, the negative (brown) streak centred at approximately $(z,y) = (\pi/4,
-0.5)$ arises due to uplift of negative streamwise velocity from the base flow
near the lower wall. Likewise, the other streaks emerge at regions of inflow and
outflow from the boundaries, with the strongest in-plane advection accounting
for the largest streaks, centred along the line $z = 0 \equiv \pi$. In the
middle panel at $x = \pi / 2$, there is transport between the upper and lower
half-channels due to vortices centred at $(\pi/4, 0)$ and $(3\pi/4, 0)$ and some
of the streaks have merged into larger structures. 
%
%We note that this indicates marked streamwise variation within the solution at
%this Reynolds number.
%
The final panel at $x = \pi = L_x/2$ is dictated by the imposed symmetries
(specifically $\shiftrefl$ and $\spanrefl$), which constrain it to be equal to
an $L_z/2$-spanwise shift of the $x=0$ slice. Likewise, the remainder of the the
computational domain, $\pi < x \leq 2\pi$, is determined this way.  Slices at
intermediate streamwise co-ordinates gradually interpolate between the panels
shown and all the states considered herein possess either a similar, or lower
level of streamwise variation.  Therefore, in the following we only use slices
at $x = 0$ and $\pi / 2$ to visualise states or, more commonly take a streamwise
average.

To examine how $EQ_7$ (and where possible, $EQ_8$) is affected by varying the
Prandtl number, a pseudo-arclength continuation algorithm was used to converge
states at fixed $\Rey = 400$ and varying $\Pran$ and $\Ri_b$.  Starting with
initial parameters $\Pran = 1$, $\Ri_b = 0$, continuation was first performed on
$\Pran$ in both directions to obtain solutions with a wide spread of Prandtl
numbers ranging from $\Pran = 10^{-4}$ to $\Pran = 500$.
%States were then continued in $\Ri_b$ to trace out the solution branches as a
%function of globally imposed stratification.
The character of these states depends strongly on whether the Prandtl number is
significantly less or greater than unity and we divide our study along these
lines.

\begin{figure}
	\centering
%	\input{curves.pgf}
	%\importpgf{.}{curves.pgf}
	\includegraphics{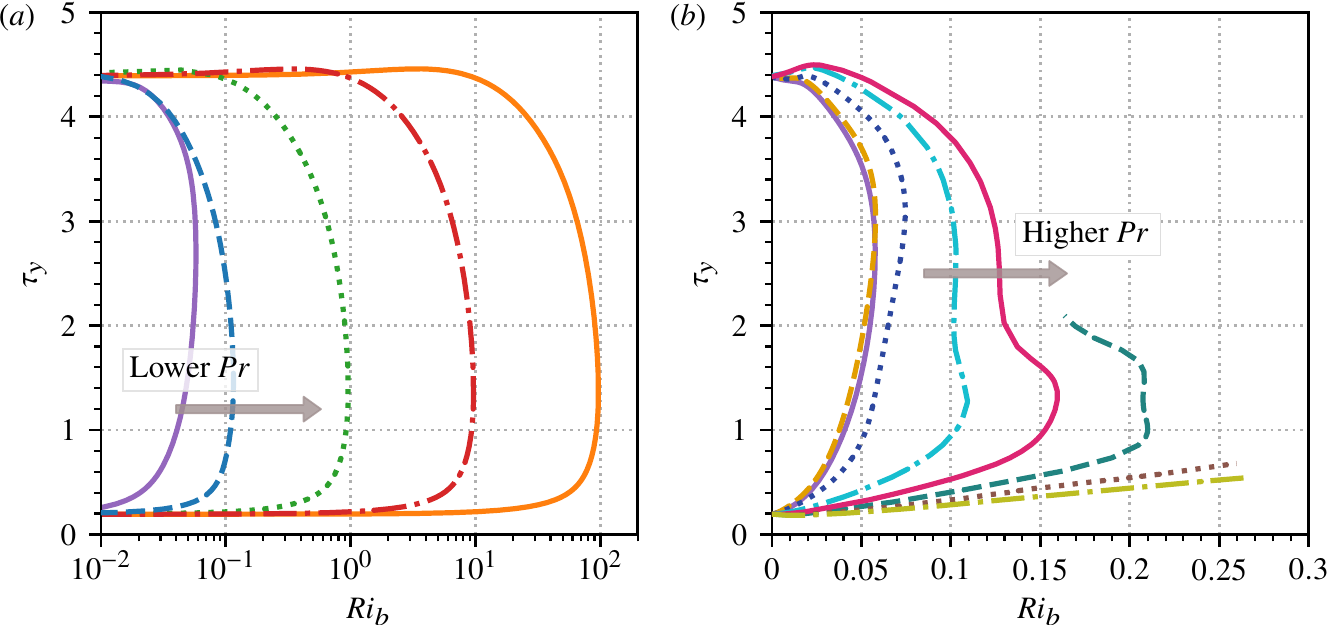}
	\caption{
    Parameter continuation curves for the $EQ_7$ (lower) / $EQ_8$ (upper) states
    in terms of the bulk Richardson number and mean wall stress $\tau_y$ for
    different Prandtl numbers.  
    (\emph{a})~Low-$\Pran$ states: $\Pran = 10^0$ (purple solid), $\Pran =
    10^{-1}$ (blue dashed), $\Pran = 10^{-2}$ (green dotted), $\Pran = 10^{-3}$
    (red dash-dot) and $\Pran = 10^{-4}$ (orange solid).  The bifurcation point
    of the $\Pran = 10^{-4}$ curve is at $\Ri_b \approx 97$.
    (\emph{b})~Higher-$\Pran$ states: $\Pran = 1$ (purple solid), $2$ (yellow
    dashed), $7$ (navy dotted), $20$ (cyan dash-dot), $40$ (magenta solid), $70$
    (teal dashed), $120$ (brown dotted), $200$ (olive dash-dot).
	}
	\label{fig:curves}
\end{figure}

Figure~\ref{fig:curves} shows solution branches of the $EQ_7$ and $EQ_8$ states,
obtained by continuing our $\Pran = 10^{-4}$ to $200$ states in $\Ri_b$.
%
% Fig 2
%
Part~(\emph{a}) plots the continuation curves for $\Pran \leq 1$. All five
curves are similar, possessing the familiar shape of a saddle-node bifurcation.
Starting from $\Ri_b = 0$, both lower and upper branches enter an initial regime
(present, but not shown for $\Pran = 1$) where their mean wall stress is
unaffected by the presence of global stratification. This is followed by a
transition between solution branches where the stress changes comparatively
rapidly. The range of admissible bulk Richardson numbers increases dramatically
as $\Pran$ decreases, suggesting that in the limit of low $\Pran$, ECS are
insensitive to stratification and may persist up to arbitrarily high $\Ri_b$ as
$\Pran$ is reduced.

%
%The remainder of this study is an analysis of the structure and physics of the
%solutions described by these continuation curves.

Figure~\ref{fig:curves}(\emph{b}) shows the $\Pran \geq 1$ continuation curves.
The $\Pran = 1$, $2$ and $7$ curves have essentially the same shape as the
low-$\Pran$ curves. At higher $\Pran$, the solution branches follow the opposite
trend to the low-$\Pran$ case, reaching progressively higher $\Ri_b$ as the
Prandtl number \emph{increases}. 
Another trend to note is that the upper branches reach increasingly higher mean
wall stresses than $EQ_8$ itself.
The $\Pran \geq 40$ curves required high numerical resolution to be continued
reliably, for reasons which shall become clear in \S\ref{sec:highpr}.  The three
highest $\Pran$ curves, which terminate before reaching $EQ_8$, were continued
as far as available computational resources permitted, the final two not
reaching their saddle-node bifurcations. Nevertheless, the $\Pran = 70$, $120$
and $200$ curves persist up to at least $\Ri_b = 0.21$, $0.26$ and $0.27$
respectively.  This trend along the lower branch suggests that $EQ_7$ might
admit arbitrarily large global stratification in the high-$\Pran$ limit. We now
analyse the solutions of Fig.~\ref{fig:curves} in the two defined limits.

%-------------------------------------------------------------------------
\subsection{Low Prandtl number ($\Pran \ll 1$)}\label{sec:lowpr}%
At low $\Pran$, density transport in the fluid is diffusion dominated and as a
result $\rho$ should not develop small scales or deviate far from the basic
state. To reflect at least the latter, we  consider the expansion $\rho(x,y,z) =
-y + \varepsilon\rho_1(x, y, z) + \ldots$, where $\varepsilon$ is a small
parameter to be determined, with $\rho_1=O(\Pran^0)$. This is certainly true at
either end of the continuation curves in Fig.~\ref{fig:curves}(\emph{a}) where
$\Ri_b \rightarrow 0$, but will be found to be true over the whole curve.
Working with the perturbations to the base state $\hat{u}\coloneqq u - y$,
$\hat{v}\coloneqq v$, $\hat{w}\coloneqq w$, $\hat{p}\coloneqq p - \Ri_b y^2 / 2$
and $\hat{\rho}\coloneqq \rho + y= \eps \rho_1+\ldots$, the steady form of 
Eq.~\eqref{eq:stratfull} becomes
\begin{equation}
    \varepsilon \left(y+\hat{u}\right) \frac{\partial \rho_1}{\partial x} + 
    \hat{v}\left( -1 + \varepsilon \frac{\partial \rho_1}{\partial y} \right)
    + \varepsilon \hat{w} \frac{\partial \rho_1}{\partial z}
    =
    \frac{\varepsilon}{\Rey\Pran} \nabla^2 \rho_1,
    \label{eq:density_pert_asym1}
\end{equation}
after neglecting $O(\eps^2)$ terms. We now determine
the leading-order velocity fields %-- particularly for $\hat{v}$ --  
%to use 
in Eq.~(\ref{eq:density_pert_asym1}) from the asymptotic VWI structure
in Eqs.~\eqref{eq:vwi1}--\eqref{eq:vwi4}.  The cross-stream velocity $\hat{v}$ is
largest for the wave field in the critical layer, but because of the enhanced
gradients of $O(\Rey^{1/3})$ across this region, this flow actually drives a
smaller density field of $O(\eps \Rey^{-1/2} )$ than the streamwise rolls
outside the critical layer, which drive an $O(\eps)$ density field. Therefore,
the leading order physics is away from the critical layer, so that
$\hat{v}=\bar{v}(y,z)/\Rey$ and $\hat{w}=\bar{w}(y,z)/\Rey$ and
Eq.~(\ref{eq:density_pert_asym1}) can be rewritten as
\begin{equation}
    \bar{v}\left( -1 + \varepsilon \frac{\partial \rho_1}{\partial y} \right)
    + \varepsilon \bar{w} \frac{\partial \rho_1}{\partial z}
    =
    \frac{\varepsilon}{\Pran} \nabla^2 \rho_1,
    \label{eq:density_pert_asym}
\end{equation}
where, because the driving term is streamwise-independent, $\rho_1=\rho_1(y,z)$
and the streamwise advection term drops. This then leads to $\varepsilon = \Pran
\ll 1$ and the leading equation 
\begin{equation}
 -\bar{v} =\nabla^2 \rho_1.
\label{eq:lowpr_leadingstrat}
\end{equation}
Inverting this relationship, the leading density perturbation
$\hat{\rho}=-\Rey\Pran\Delta^{-1} \hat{v}$ (where $\Delta\coloneqq\nabla^2$) can
then be eliminated from the Navier-Stokes equations to leave the low-$\Pran$
equations \citep[or low-P\'{e}clet number equations, see][]{Lignieres1999}:
\begin{subequations}
\begin{align}
  \left(y + \hat{u}\right) \hat{u}_x + 
 \hat{v} \left( 1 + \hat{u}_y \right) + 
  \hat{w} \hat{u}_z + \hat{p}_x &= \Rey^{-1} \nabla^2\hat{u}, \\
  \left(y + \hat{u}\right) \hat{v}_x + \hat{v}\hat{v}_y + \hat{w}\hat{v}_z
  + \hat{p}_y &= \Rey^{-1} \left[ \nabla^2 + \Ri_b\Pran \Rey^2 \,
  \Delta^{-1}\right]\hat{v},\label{eq:vmompert}\\
  \left(y + \hat{u} \right) \hat{w}_x + \hat{v}\hat{w}_y +
   \hat{w}\hat{w}_z + \hat{p}_z &= \Rey^{-1}\nabla^2 \hat{w}, \\
  \hat{u}_x + \hat{v}_y + \hat{w}_z &= 0.
%  \left(y + \hat{u}\right) \hat{\rho}_x + 
%   \hat{v}\left(-1 + \hat{\rho}_y \right) + \hat{w}\hat{\rho}_z &=
%  \Rey^{-1}\Pran^{-1} \nabla^2 \hat{\rho}.\label{eq:densitypert}
\end{align}
\end{subequations}
This makes it clear that solutions can only depend on the the Richardson number
$\Ri_b$ and the Prandtl number $\Pran$ in the combination $\Ri_b \Pran$.
Furthermore, the cross-stream momentum equation, Eq.~(\ref{eq:vmompert}), also implies
that there exists a region of weakly stratified (small $\Ri_b$) flow where the
Navier-Stokes and the density equations remain effectively uncoupled until
$\Ri_b=O(\Pran^{-1}\Rey^{-2})$.  These are the flat regions of both the upper and
lower solution branches in Fig.~\ref{fig:curves}(\emph{a}), where increasing
$\Ri_b$ has no effect on $\tau_y$. As $\Pran \rightarrow 0$, these regions
extend to $\Ri_b \sim O(1/\Pran)$.

\begin{figure}
%	\input{lowprscalings.pgf}
	%\importpgf{.}{lowprscalings.pgf}
	\includegraphics{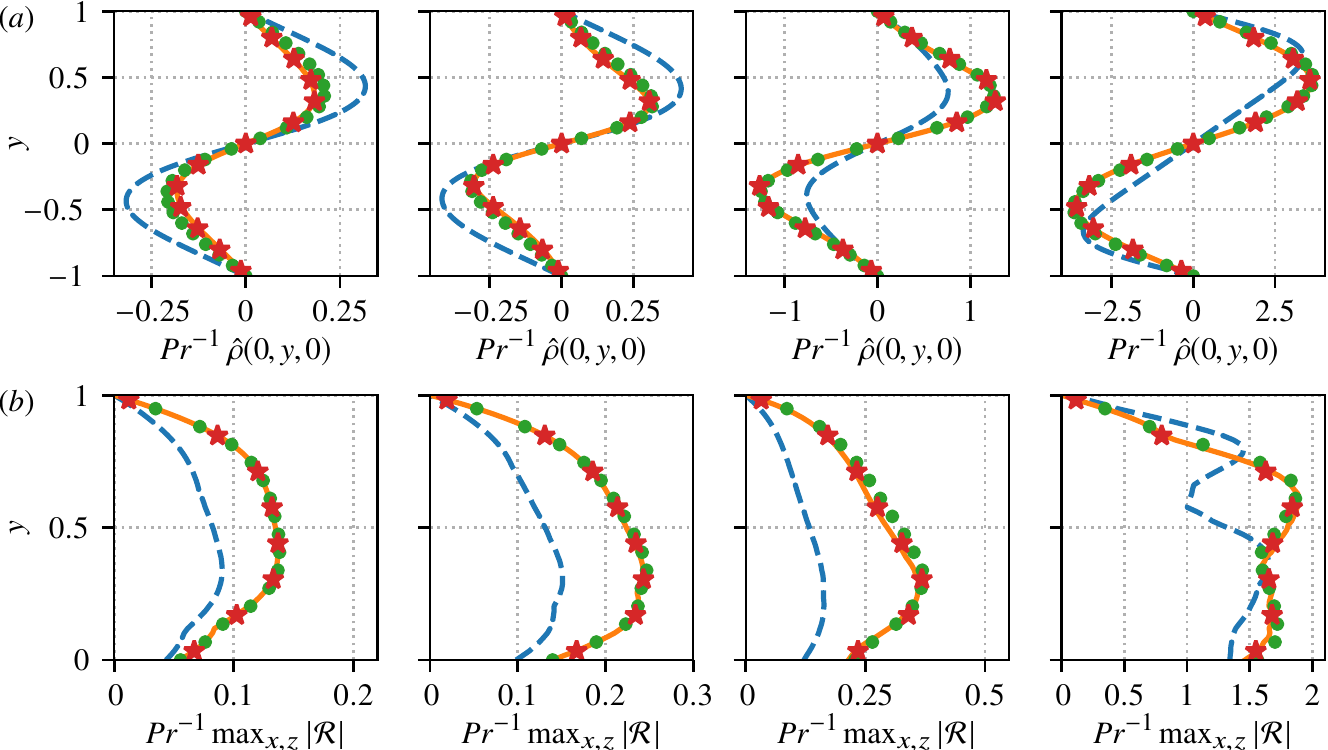}
    \caption{
        Verification of the $O(\Pran)$ scalings predicted for low Prandtl
        number ECS.  In all plots the Prandtl numbers are $\Pran = 10^{-1}$
        (blue dashed), $\Pran = 10^{-2}$ (green circles) $\Pran = 10^{-3}$ (red
        stars) and $\Pran = 10^{-4}$ (orange solid).  Across each column
        $\Pran\Ri_b$ is fixed at a given point along the solution curves plotted
        in Fig.~\ref{fig:lowpr}(\emph{a}). From left to right: $\Pran\Ri_b = 0$,
        $0.005$ (lower branch), $0.097$ (saddle-node) and $0.005$ (upper
        branch).
        (\emph{a})~Cross-channel
        profiles of $\hat{\rho}$, for $x, z = 0$, 
        rescaled by $\Pran^{-1}$.
        (\emph{b})~Absolute maximum values along streamwise and spanwise
        directions of %of the residuals of Eq.~\eqref{eq:lowpr_leadingstrat},
        $\mathcal{R} \coloneqq \hat{v} + \nabla^2\hat{\rho} / (\Rey\Pran)$, 
        rescaled by $\Pran^{-1}$.
    }
	\label{fig:lowprscaling}
\end{figure}

This scaling result, $\Ri_b = O(\Pran^{-1}\Rey^{-2})$, equivalent to the
Rayleigh number $\Ra\coloneqq-\Ri_b\Pran\Rey^2$ being $O(1)$, was  first
observed numerically in the $\Pran = 1$ case by~\citet{Eaves2015} and then in
the analysis of~\citet{Deguchi2017} and \cite{Olvera2017}. There, at least for
$\Pran =O(1)$, once the stratification strength increases beyond this scaling,
new dynamic balances emerge [e.g.\ `regime 2' in \cite{Olvera2017}, ultimately
followed by the `unit Reynolds number Navier-Stokes' (UNS) regime of
\citet{Deguchi2017} or `regime 3' of \citet{Olvera2017}]. Here however, in the
$\Pran \rightarrow 0$ limit this is not the case.  In
Fig.~\ref{fig:lowprscaling}(\emph{a}), we demonstrate that the $O(\Pran)$
scaling of the density perturbation persists along the \emph{whole} length of
the solution curves.
The leftmost figure verifies the situation for $EQ_7$ at $\Ri_b = 0$ (where the
stratification is still present but the buoyancy force is switched off),
plotting $\Pran^{-1}\hat{\rho}$ as a function of $y$ for $x, z = 0$.  These
slices, taken from solutions at $\Pran = 10^{-1}, 10^{-2}, 10^{-3}$ and
$10^{-4}$, collapse neatly onto an asymptotic profile for $\Pran \gtrsim
10^{-2}$.  The subsequent plots confirm that the same holds for  $\Ri_b > 0$
solutions using the same Prandtl numbers, at various points along the solution
branches, now keeping the size of the buoyancy forcing constant.  This property
is not specific to our choice of $x$, $z = 0$. We have checked that the scalings
are observed for profiles with $x = 0$, $\pi/4$, $\pi/2$, $3\pi/4$ and $z
= 0$, $\pi / 2$. 

\begin{figure}
%	\input{lowpr.pgf}
	%\importpgf{.}{lowpr.pgf}
	\includegraphics{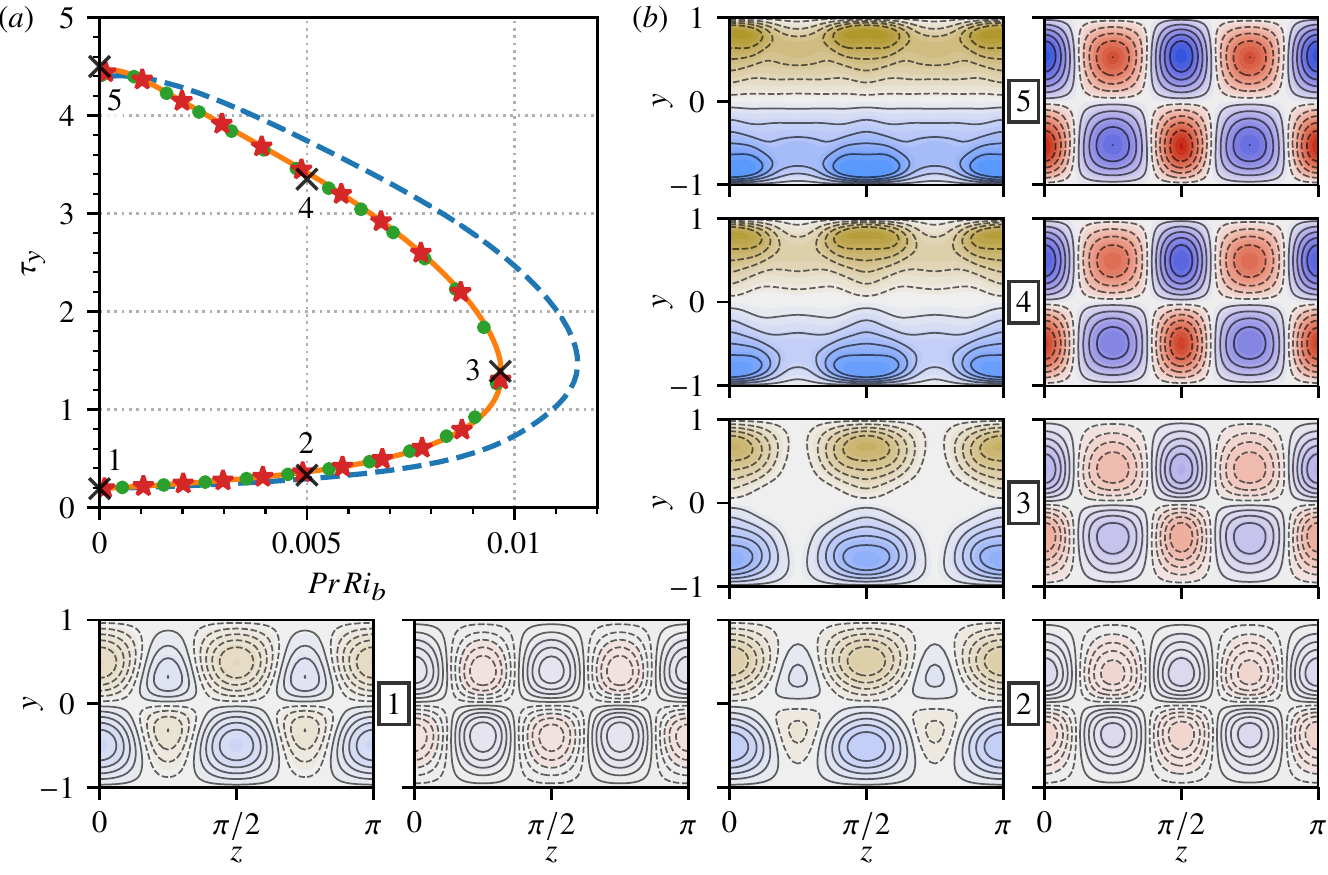}
    \caption{
		Self-similar solution branches of $EQ_7$ (lower) / $EQ_8$ (upper) in
		the $\Pran\to 0$ limit.
        (\emph{a})~Parameter continuation curves in bulk Richardson
        number as a function of mean wall stress. The horizontal axis has been
        rescaled by $\Pran$ causing the curves to collapse.
        The Prandtl numbers match those in Fig.~\ref{fig:lowprscaling}.
        (\emph{b})~Pairs of contour plots along the $\Pran = 10^{-4}$ branch
        (orange solid), numbered according to the labelled points in part
        (\emph{a}).  Lefthand plots show the streamwise average of $\hat{u}$,
        shaded between $-0.8$ (brown) and $0.8$ (blue).  Righthand plots show
        the streamwise average of the $\hat{\rho}$ field, shaded between
        $-2.3\times 10^{-4}$ (red) and $2.3\times 10^{-4}$ (blue). In all plots,
        the middle contour interval is white and centred at zero; black contour
        lines are overlaid to show the structure of solutions, using $11$
        equally spaced intervals, between $\pm \max_{y,z} |\bar{u}-y|$ or $\pm
        \max_{y,z}|\bar{\rho}|$ for each field, as appropriate.  
        Dashed lines
        are negative contours and solid lines are positive. 
        Anti-clockwise from
        bottom left: $\Pran\Ri_b= 0$ ($EQ_7$, point 1), $0.005$ (point 2),
        $0.097$ (saddle-node, point 3), $0.005$ (point 4), $0$ ($EQ_8$, point
        5).
	}
	\label{fig:lowpr}
\end{figure}

In Fig.~\ref{fig:lowprscaling}(\emph{b}), we plot $\max_{x,z} | \hat{v}+\nabla^2
\hat{\rho} / (\Rey\Pran)|$ at the same points along the solution curves as in
Fig.~\ref{fig:lowprscaling}(\emph{a}). These curves collapse upon rescaling by
$\Pran^{-1}$, confirming that Eq.~\eqref{eq:lowpr_leadingstrat} is obeyed up to
an $O(\eps)$ error term, as implied by Eq.~\eqref{eq:density_pert_asym}.  The
conclusion is then that solutions should be self-similar in the limit of low
$\Pran$ along the whole solution curve.  Figure~\ref{fig:lowpr}(\emph{a})
replots the parameter continuation curves in Fig.~\ref{fig:curves}(\emph{a})
against $\Ri_b \Pran$, confirming the collapse onto a single asymptotic branch.
In particular, this implies that the maximum bulk Richardson number for these
solutions is $\Ri^{m}_b \approx  0.01/\Pran$ at $\Rey=400$ as $\Pran \to 0$.

\begin{figure}
%    \input{lowpr_v.pgf}
	%\importpgf{.}{lowpr_v.pgf}
	\includegraphics{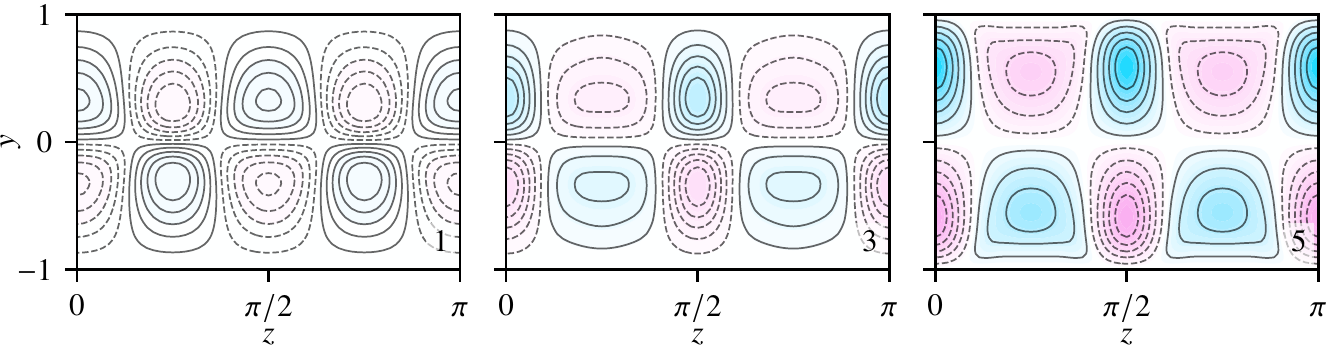}
    \caption{%
        Contour plots of $\bar{v}(y,z)$ along the $\Pran = 10^{-4}$ solution
        branch, at the numbered points indicated in
        Fig.~\ref{fig:lowpr}(\emph{a}):
        from left to right, the plots correspond to the first ($EQ_7$), third
        (saddle-node) and fifth ($EQ_8$) solution pairs in
        Fig.~\ref{fig:lowpr}(\emph{b}). 
        The contours are shaded between $-0.19$ (pink) and $0.19$ (cyan); black
        contour lines are overlaid using $11$ equally spaced intervals, between
        $\pm \max_{y,z} |\bar{v}|$, which equals $0.014$, $0.076$
        and $0.19$ respectively.
        Dashed contours indicate $\bar{v} < 0$, solid contours $\bar{v} > 0$.
	}
	\label{fig:lowprv}
\end{figure}

\begin{figure}
	%\input{lowpr_lam.pgf}
	%\importpgf{.}{lowpr_lam.pgf}
	\includegraphics{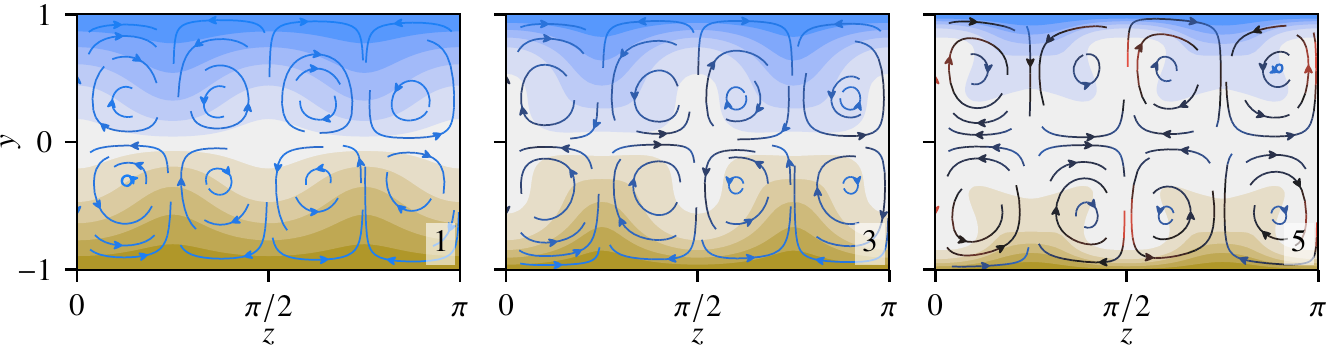}
    \caption{
    Contour plots of $\bar{u}(y,z)$ for $\Pran = 10^{-4}$ solutions at the
    numbered points indicated on the solid orange Fig.~\ref{fig:lowpr}(\emph{a})
    curve.  From left to right, the plots correspond to the first ($EQ_7$),
    third (saddle-node) and fifth ($EQ_8$) plotted solutions in
    Fig.~\ref{fig:lowpr}(\emph{b}).
    Overlaid are $(\bar{v},\bar{w})$-streamlines,
    coloured with a linear gradient from blue to black to red, according to
    $|(\bar{v},\bar{w})|$, which lies in the interval $[0, 0.19)$.
	}
	\label{fig:lowpr_lam}
\end{figure}

In Fig.~\ref{fig:lowpr}(\emph{b}) we plot contour slices through the $yz$-plane
along the $\Pran = 10^{-4}$ curve. The contour plots are organised in numbered
side-by-side pairs, showing streamwise averages of $\hat{u}$ on the left and
$\hat{\rho}$ on the right.  Along the lower branch (plots $1$--$3$) we see that
the streaks centred at $z = \pi / 4$ and $3\pi / 4$ shrink as $\Ri_b$ increases.
At the saddle-node (plot 4) they have disappeared completely and the mean
$\hat{u}$ field becomes separated into negative (upper) and positive (lower)
halves.
This can be traced to the $\hat{v}$ field, which feeds these streaks by
advecting the base shear; its streamwise average is plotted in
Fig.~\ref{fig:lowprv}. 
(For reference, the corresponding roll streamlines are included in
Fig.~\ref{fig:lowpr_lam}.)
As the state transitions from the less energetic $EQ_7$ state toward the more
energetic $EQ_8$, the overall strength of the $\hat{v}$ field increases (as do
the magnitudes of $\hat{u}$ and $\hat{\rho}$, which are driven by $\hat{v}$).
However, the increased buoyancy force from point 1 to point 3 penalises upward
motion wherever $\hat{\rho} > 0$ and downward motion where $\hat{\rho} < 0$.
Consequently, there is a redistribution of the $\hat{v}$ field along the lower
branch, diminishing in regions where the signs of $\hat{v}$ and $\hat{\rho}$
match (outflow from the walls) and concentrating where they differ (inflow to the
walls).  This leads to the (inflow) streaks centred at $z = 0 \equiv \pi$ and
$\pi / 2$ becoming increasingly dominant.
This division persists along much of the upper branch, connecting through to the $EQ_8$ solution where
the unevenly distributed wall-normal velocity field and corresponding streak
structure exist in the unstratified setting.

%From the saddle-node onwards, these trends continue, despite the solutions being
%unable to access higher $\Ri_b$. Between the saddle-node and the upper branch
%halfway point [third and fourth pairs in Fig.~\ref{fig:lowpr}(\emph{b})],
%$\hat{\rho}$ increases, as does $\Ri_b \max_{x,y,z} | \hat{\rho} |$ \JL{double
%check}, which is a good proxy for the typical strength of buoyancy forces.
%Correspondingly, inflow to the walls increases and outflow is suppressed; the
%dominant streaks are pushed in toward the boundary and lose some spatial
%coherence in the spanwise direction. Moreover, their magnitude increases leading
%to substantially higher wall stresses. Finally, between the midpoint of the
%upper branch and $EQ_8$ [fourth and final pairs in
%Fig.~\ref{fig:lowpr}(\emph{b})], the strength of buoyancy forcing decreases to
%zero. Nevertheless, the magnitude of the density perturbation continues to
%increase and so does $\hat{v}$. Therefore, $\hat{u}$ continues to strengthen and
%spanwise-delocalise.

The qualitative effect of these trends along the solution branches on the
streamwise transport is shown in Figure~\ref{fig:lowpr_lam}, which shows the
$\bar{u}$ field for $EQ_7$ at $\Ri_b = 0$ (left plot), at the saddle-node
(middle plot) and $EQ_8$ at $\Ri_b = 0$ (right plot).
This shows that by the end of the lower branch, $\hat{u}$ is large enough to
cancel the base laminar shear flow solution in some regions (i.e.\ $u=y+\hat{u}=0$).  The overlaid
streamlines highlight the underlying roll advection, its increasing significance
and the uneven distribution of inflow versus outflow to the walls.  However, it
is important to note that while streamwise-averaged quantities indicate overall
trends, the streamwise dependence observed in Fig.~\ref{fig:eq7} is present and
persists along the solution curves. Specifically, it is not the case that $u$ is
homogenised throughout the channel interior along the upper branch, even though this is
true in a streamwise-averaged sense.

%----------------------------------------------------------------------------
%
%  3.2    Pr >> 1
%
%----------------------------------------------------------------------------

\subsection{High Prandtl number ($\Pran \gg 1$)}
\label{sec:highpr}
\begin{figure}%
	%\input{ripos_ex.pgf}%
	%\importpgf{.}{ripos_ex.pgf}%
	\includegraphics{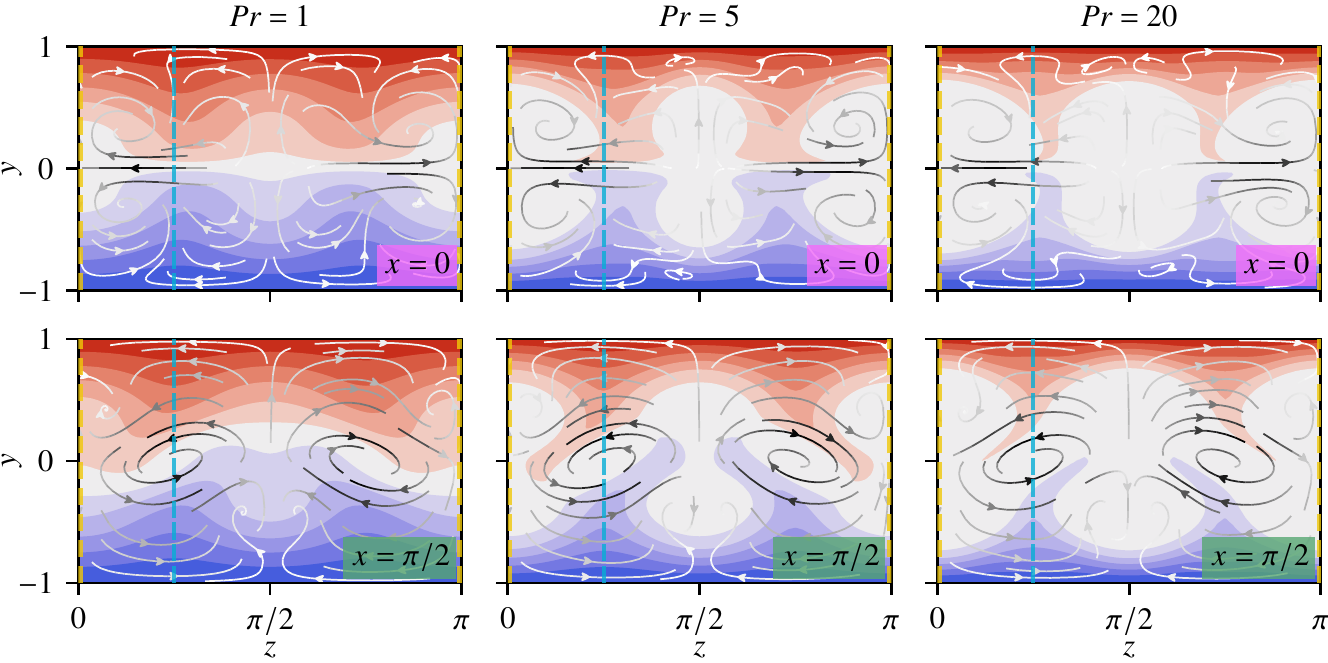}
    \caption{%
        Contour plots of $\rho(x, y, z)$, through $x=0$ (top row) and $x =
        \pi/2$ (bottom row) of the $EQ_7$ state at $\Pran = 1$, $5$ and $20$
        (left to right) and $\Ri_b = 0.01$.  Contours are equally spaced from
        $-1$ (red) to $1$ (blue) with a width of $2/11$, ensuring that the
        middle band (white) is centred at zero.  Overlaid are
        $(v,w)$-streamlines, coloured with a linear gradient from white to black
        according to the magnitude of $(v, w)$ across all six plots, which lies
        in the interval $[0, 0.154)$.  The coloured dashed lines through $z=0$
        and $\pi/4$ indicate the locations of the contour slices plotted in
        Fig.~\ref{fig:highpr_example_xy}.
    }%
    \label{fig:highpr_example}%
\end{figure}

Just as in the case of low Prandtl number, solution curves calculated with large
$\Pran$  access higher and higher bulk Richardson numbers as $\Pran$ increases.
However, the mechanism which allows this is quite different, since $\hat{\rho}$
is no longer a weak perturbation of the uniform laminar density profile.  In
Fig.~\ref{fig:highpr_example} we plot contour slices through the $yz$-plane at
$x = 0$ (top row) and $\pi/2$ (bottom row) of the full density field $\rho$ for
$\Pran = 1$, $5$ and $20$.
The bulk Richardson number is $0.01$ and the state is the lower branch $EQ_7$
solution. Starting at $\Pran = 1$ (leftmost column), the density field for $x=0$
is a spanwise-wavy modulation of the base profile.  Centred at $z=0$, there is a
well-mixed region, where $\rho\approx 0$, sandwiched between two regions of
strong spanwise inflow along the centreline $y=0$. In the $\pi/2$ slice, in-plane
advection is dominated by vortices centred at $(z, y) = (\pi/4, 0)$ and
$(3\pi/4,0)$, inducing a global spanwise wave.
At $\Pran = 5$ (middle column) the spanwise waviness in both $x$ slices has
become much more exaggerated, with peaks of extremal density advected towards
$y=0$ and corresponding troughs sent out towards the walls. In between these
features, regions of approximately zero density are now well established. By
$\Pran = 20$ (rightmost column), the spanwise-wavy structures have become
spanwise localised `finger'-like incursions into an otherwise uniformly dense
and neutrally buoyant channel interior.  Moreover, the regions in the vicinity
of the walls have become highly stratified.

\begin{figure}%
	%\input{ripos_xy.pgf}
	%\importpgf{.}{ripos_xy.pgf}%
	\includegraphics{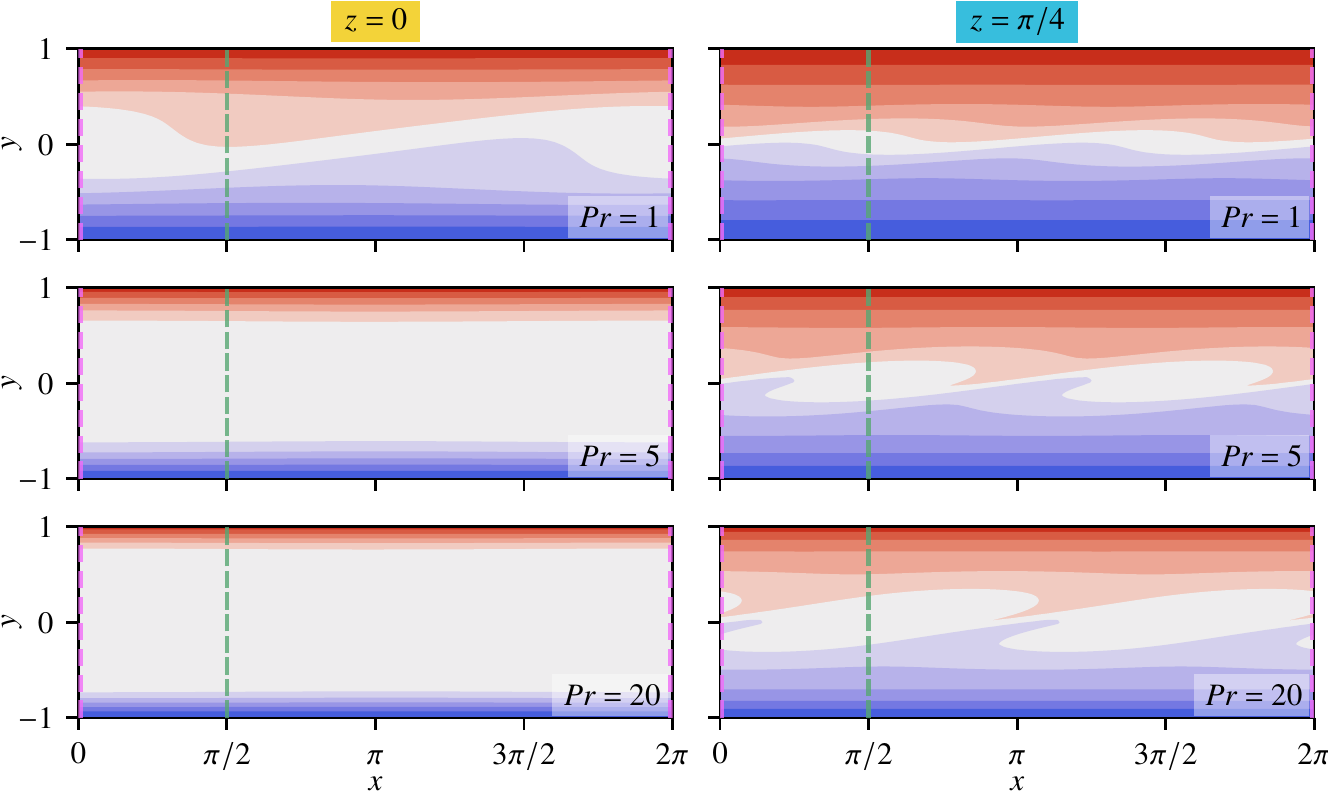}
    \caption{%
    Contour slices through $xy$-planes, of $\rho(x, y, z)$, at $z=0$ (left
    column) and $z = \pi / 4$ (right column) of the $EQ_7$ state at at $\Pran =
    1$, $5$ and $20$ (top to bottom) and $\Ri_b = 0.01$.  The solutions match
    those depicted in Fig.~\ref{fig:highpr_example}, as do the contour
    intervals.
    The coloured dashed lines at $x = 0$ and $\pi / 2$ mark the locations of the
    contour slices depicted in Fig.~\ref{fig:highpr_example}.
    }%
    \label{fig:highpr_example_xy}%
\end{figure}

Figure~\ref{fig:highpr_example_xy} shows contour slices in the $xy$-plane
using the same solutions as Fig.~\ref{fig:highpr_example}. 
In this case, the Prandtl number increases moving down each column. The left
column is a slice through $z = 0$; we see that the interior in this region is
already homogenised by $\Pran = 5$ across the full length of the channel. The
righthand column is a slice through $z = \pi / 4$. Even in this region, by
$\Pran = 20$, the interior appears to be homogenising, despite the
aforementioned inflow of density from the walls.

\begin{figure}
	%\input{bldevel.pgf}
	%\importpgf{.}{bldevel.pgf}%
	\includegraphics{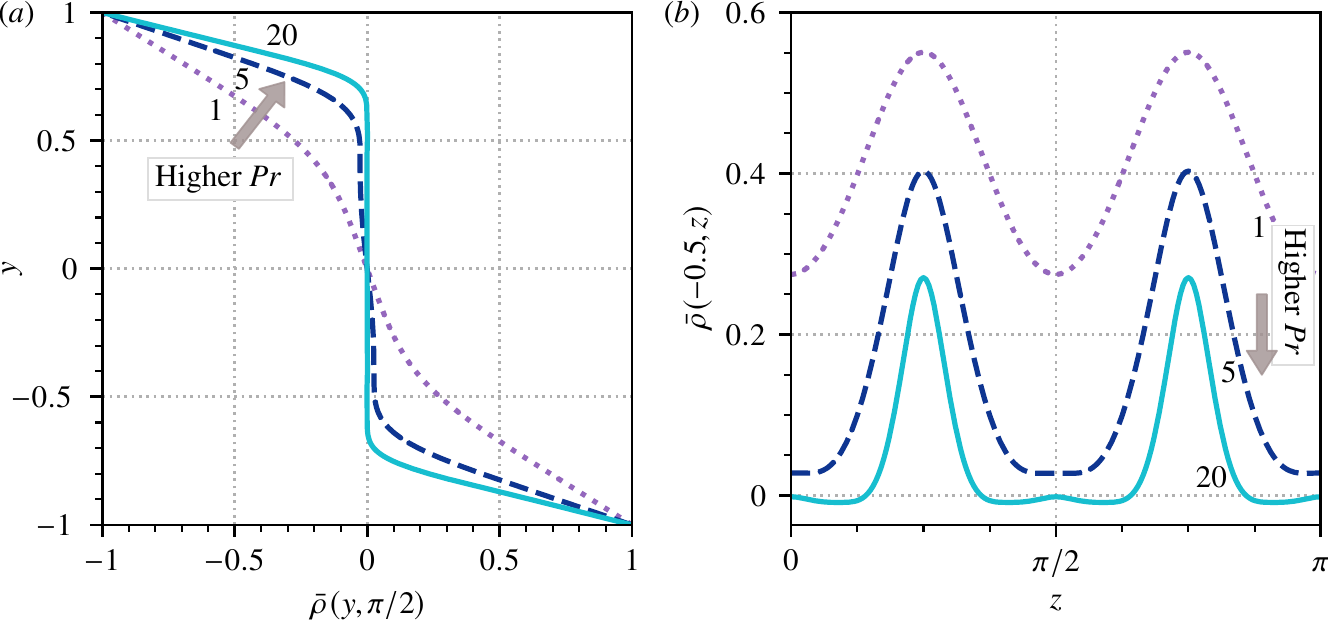}
    \caption{%
    Localisation of stratification as $\Pran$ increases. The data are from
    the $\Ri_b = 0.01$ solutions depicted in the contour plots of
    Figs.~\ref{fig:highpr_example} and~\ref{fig:highpr_example_xy}. The Prandtl
    numbers are: $\Pran = 1$ (purple dotted), $\Pran = 5$ (navy dashed), $\Pran
    = 20$ (cyan solid).  (\emph{a})~Vertical profiles of $\bar{\rho}(y,z)$ for
    fixed $z = \pi / 2$, showing the development of boundary layers at the
    walls.
    (\emph{b})~Horizontal slices of $\bar{\rho}(y,z)$ for fixed $y = -0.5$,
    showing the shrinking width and magnitude of the regions of extremal density
    uplifted from the boundaries.
	}
	\label{fig:bldevel}
\end{figure}

At the channel walls, the homogenising density profile must connect to
$\rho(x,\pm 1,z) = \mp
1$. Consequently, as the well-mixed interior region expands, we observe the
profile at the walls steepening, developing into separate boundary layers.
This is demonstrated in Fig.~\ref{fig:bldevel}(\emph{a}), in which we plot the
streamwise-averaged density $\bar{\rho}$ through the line $z = \pi / 2$.
At $\Pran = 1$, $\bar{\rho}$ has an almost linear profile. By $\Pran = 20$, the
profile is essentially flat in the interior before becoming highly stratified at
the walls. These profiles are reminiscent of turbulent mean temperature
profiles of turbulent passively transported scalars simulated by
\cite{Papavassiliou1997}, of fully stratified turbulence
simulations~\citep{Zhou2017} and of strongly stratified layers
observed in geophysical flows~\citep{Turner1973}. See \S\ref{sec:discussion}
for further discussion.

Simultaneously, the finger-like incursions, or `density fingers', seen in Fig.~\ref{fig:highpr_example}
start to diminish as the channel becomes
increasingly well-mixed. Figure~\ref{fig:bldevel}(\emph{b}) shows horizontal
slices of $\bar{\rho}$ at $y = -0.5$. We see that these
structures progressively narrow as $\Pran$ increases. Moreover, from $\Pran = 5$
to $\Pran = 20$ their `height' ($\max_z\bar{\rho}$) and `prominence'
($\max_z\bar{\rho}-\min_z\bar{\rho}$) diminishes. These observations are consistent with the
emerging homogenisation of the channel interior region through the plane $z =
\pi / 4$, seen in Fig.~\ref{fig:highpr_example_xy}.

It is the diminishing length scales described above that make obtaining
solutions at high Prandtl number numerically expensive, ultimately forcing us to
cut short some of the continuation curves in Fig.~\ref{fig:curves}, since states
demand increasingly high wall-normal and (to a lesser extent) spanwise
resolution as $\Pran$ increases. Accurately converging states in situations
where structures become spatially localised is an important future challenge
\citep[see also][]{Olvera2017}. In the following section we obtain higher
Prandtl numbers by moving to a regime where we need only solve for the density
field (see Appendix~\ref{appendix:numerics}.)

%
%
% 3.2.1
%
%
\subsubsection{Passive scalar limit ($\Ri_b \to 0$)}
\label{sec:ri0}
\begin{figure}
	%\includegraphics[width=5.3in]{figures/rhobaryz_prup.jpg}
	%\input{rhobar_prup.pgf}
	%\importpgf{.}{rhobar_prup.pgf}%
	\includegraphics{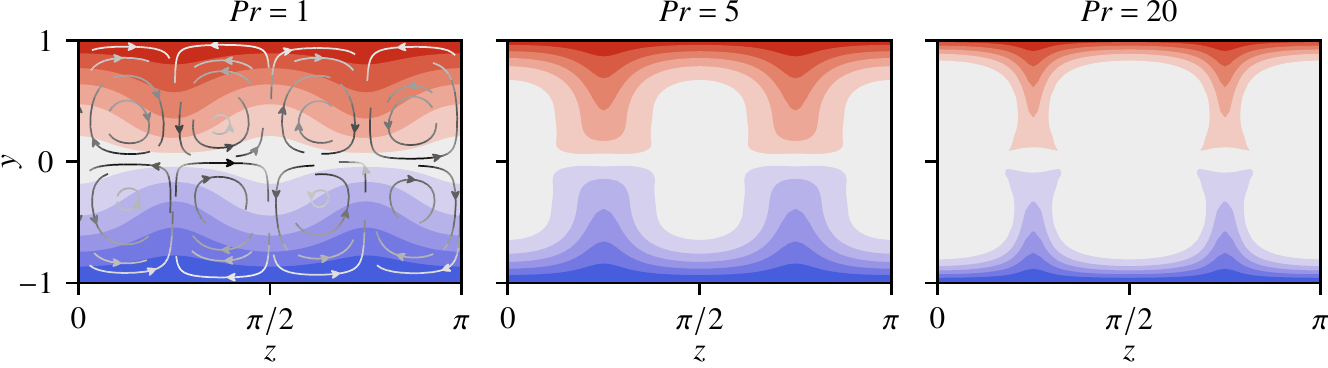}
    \caption{Contour plots of $\bar{\rho}_\mathrm{PS}(y,z)$ for increasing
    $\Pran$ in the channel cross-section.  From left to right: $\Pran = 1$,
    $\Pran = 5$ and $\Pran = 20$ are shown.  The contour intervals match those
    in Fig.~\ref{fig:highpr_example}.
    %Contours are equally spaced from $-1$ (blue) to $1$ (red) with a width of
    %$2/11$, ensuring that the middle band (white) is centred at zero.
    Overlaid on the leftmost plot are streamlines of the
    $(\bar{v},\bar{w})$-field (which is the same for all three solutions),
    coloured with a linear gradient from white to black according to the
    magnitude of $(\bar{v},\bar{w})$, which lies in the interval $[0, 0.02)$. 
	}
	\label{fig:rhobaryz_prup}
\end{figure}

The passive scalar limit, $\Ri_b \rightarrow 0$, offers key insight into these
high-$\Pran$ solutions, since in this regime the density field decouples from
the momentum equation [Eq.~\eqref{eq:momfull}]. 
We hereafter adopt the notation $\rho_\mathrm{PS}$ for the density fields in
this limit.
The uncoupling of momentum from density allows the stratification equation
%This allows the density equation
[Eq.~\eqref{eq:stratfull}] to be studied independently from the velocity field,
which is just that of the unstratified equilibrium state. Numerically, this
limit can be accessed simply by setting $\Ri_b=0$ in the nondimensionalised
system, Eqs.~(\ref{eq:momfull})--(\ref{eq:stratfull}).
Figure~\ref{fig:rhobaryz_prup} shows the streamwise-averaged passive scalar density
field, $\bar{\rho}_\mathrm{PS}$, for $EQ_7$ at three successively increasing
$\Pran$.
%
% Fig 10
%
This reveals an underlying structure of stacked rolls 
%(which become
%streamwise-invariant in the high $\Rey$ limit) 
and their advective influence on the density field. Much of the character of the
density fields described above persists in the $\Ri_b = 0$ case. In particular,
the spanwise-localised fingers remain, as does the stratified boundary layer,
which rapidly diminishes with increasing $\Pran$, leaving a largely homogeneous
interior. 

The increasing concentration of $\rho_\mathrm{PS}$ at the boundaries indicates
that high-$\Pran$ solutions must involve a separation of length scales.
Considering a boundary layer of thickness $\epsilon$ at the bottom boundary
$y=-1$, we define a scaled vertical coordinate $Y\coloneqq (y+1)/\epsilon$ and
Taylor expand the leading order (VWI) velocity field across the layer
\begin{subequations}%
\begin{align}%
    u(x, y, z) & = -1 + \epsilon u_1(x,z) Y+\ldots \label{eq:uexpan}\\
    v(x, y, z) &= \epsilon^2 \Rey^{-1} v_2(x,z) Y^2+ \ldots \label{eq:vexpan}\\
    w(x, y, z) &= \epsilon  \Rey^{-1} w_1(x,z) Y+ \ldots \label{eq:wexpan}%
\end{align}%
\end{subequations}%
using the boundary conditions ($u+1=v=w=0$ at $Y=0$) and incompressibility
($\partial v/ \partial y=0$ at $Y=0$). Since the boundaries are well away from
the critical layer at $y=0$, the flow fields (as $\Rey \rightarrow \infty$) have
the form
\begin{subequations}
\begin{align}
u_1(x,z) &=\bar{u}_1(z)+Re^{-7/6}U_1(x,z)+\dots,\\
v_2(x,z) &=\bar{v}_2(z) +Re^{-1/6}V_2(x,z)+\ldots,\\
w_1(x,z) &=\bar{w}_1(z)  +Re^{-1/6}W_1(x,z)+\ldots
\end{align}
\end{subequations}

\begin{figure}
	%\input{blscaling.pgf}
	%\importpgf{.}{blscaling.pgf}%
	\includegraphics{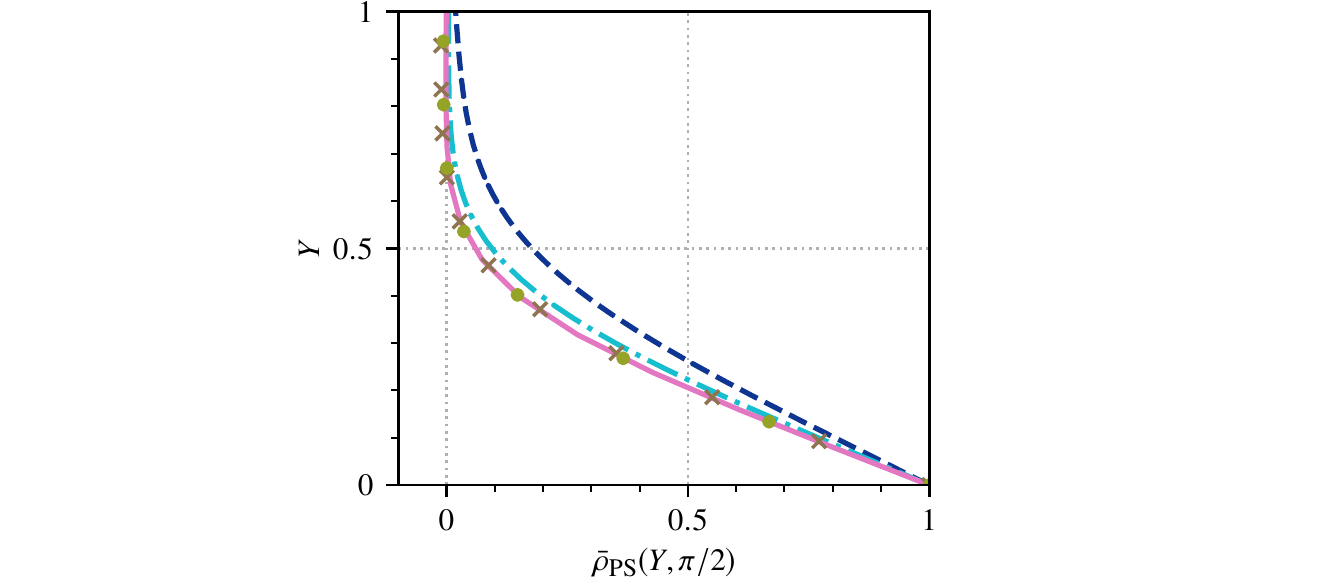}
	\caption{
        Verification of the boundary layer thickness $\epsilon = \Pran^{-1/3}$.
        Streamwise-averaged density profiles through $z = \pi/2$ are shown,
        plotted near the wall using boundary layer co-ordinates $Y \coloneqq
        \Pran^{1/3}(y+1)$.  The data are from $EQ_7$ in the passive scalar
        limit, at $\Pran = 5$ (navy dashed), $\Pran = 20$ (turquoise dash--dot),
        $\Pran = 100$ (brown crosses), $\Pran = 300$ (olive circles) and $\Pran
        = 500$ (pink solid).  The three highest Prandtl number profiles collapse
        almost exactly in the rescaled wall-normal co-ordinates.
	}
	\label{fig:blscaling}
\end{figure}

It is straightforward to argue that $\rho_\mathrm{PS}$ must be
streamwise-invariant at leading order (see Appendix~\ref{appendix:bl}) so that
an appropriate expansion for the density field inside the boundary layer is
\begin{equation}%
    \label{eq:revised_expan}%
    \rho_\mathrm{PS}(x, Y, z) = \rho_0(Y, z) + \epsilon \rho_1(x, Y, z) + \ldots,%
\end{equation}%
where the functions $\rho_0$ and $\rho_1$ are $O(\epsilon^0)$.
On substituting Eqs.~\eqref{eq:uexpan}, \eqref{eq:vexpan} and \eqref{eq:wexpan}
into Eq.~\eqref{eq:stratfull}, along with the asymptotic expansion for the
density field and dropping all $O(\epsilon^2)$ terms, we obtain the following
leading-order stratification equation:%
%In the leading order stratification equation, the wall-normal diffusion
%is balanced by $O(\epsilon)$ advective terms in all directions. 
%
\begin{equation}%
	-\epsilon \Rey\frac{\partial \rho_1}{\partial x} 
	+
	\epsilon v_2(x,z) Y^2 \frac{\partial \rho_0}{\partial Y}
	+
	\epsilon w_1(x,z) Y \frac{\partial \rho_0}{\partial z}
	=
	\frac{1}{\epsilon^2\Pran}\frac{\partial^2\rho_0}{\partial Y^2}.%
	\label{eq:blstratfull}%
\end{equation}%
Balancing  advection with diffusion then requires
\begin{equation}%
	\epsilon = \Pran^{-1/3},%
	\label{eq:epsilon}%
\end{equation}%
which is confirmed in Fig.~\ref{fig:blscaling} by plotting
$\bar{\rho}_\mathrm{PS}(Y,z)$ through the line $z =\pi/2$ (other values show similar
collapse as long as they are not within the finger-like structures). 
%
% Fig 11
%
In these co-ordinates, the density profiles collapse onto each other almost
exactly at high $\Pran$ ($\gtrsim 100$). 

Equation~\eqref{eq:blstratfull} splits into a streamwise-independent part which
defines $\rho_0$ and a streamwise-dependent part which defines $\rho_1$ given
$\rho_0$ as follows:
\begin{subequations}
\begin{align}
	\bar{v}_2(z)Y^2 \frac{\partial\rho_0}{\partial Y} + 
	\bar{w}_1(z)Y\frac{\partial\rho_0}{\partial z}  &= 
    \frac{\partial^2 \rho_0}{\partial Y^2},  \label{rho_0}\\
	\Rey^{-7/6} V_2(x,z) Y^2 \frac{\partial \rho_0}{\partial Y}
	+
	\Rey^{-7/6} W_1(x,z) Y \frac{\partial \rho_0}{\partial z}
	&=  \frac{\partial \rho_1}{\partial x}. \label{rho_1}
\end{align}
\end{subequations}
Using incompressibility, $\bar{w}_1^\prime(z) = -2\bar{v}_2(z)$, the leading
balance, Eq.~\eqref{rho_0}, becomes
\begin{equation}
    \frac{\partial^2\rho_0}{\partial Y^2} + 
	\tfrac{1}{2}\bar{w}_1^\prime(z)Y^2\frac{\partial\rho_0}{\partial Y} =
	\bar{w}_1(z)Y\frac{\partial\rho_0}{\partial z},
	\label{eq:rho0asym}
\end{equation}
with the boundary conditions $\rho_0(0, z) = 1$ and 
$\rho_0(Y,z) \rightarrow 0$ as $Y \rightarrow \infty$ to match onto a homogenised interior.
%$\lim_{Y\to\infty}\rho_0(\cdot,z) = 0$.
%

\begin{figure}
	\centering
	%\input{highprcontour.pgf}
	%\importpgf{.}{highprcontour.pgf}%
	\includegraphics{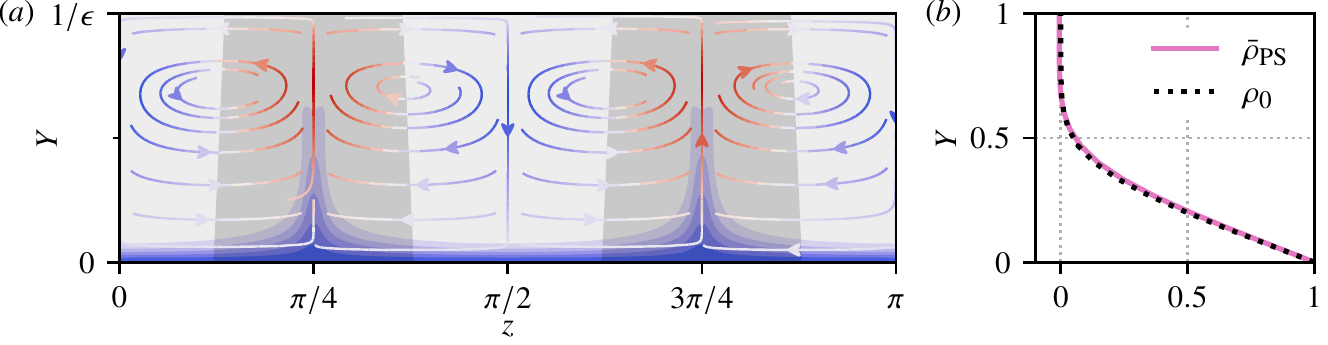}
    \caption{
    (\emph{a})~Plot of $\bar{\rho}_\mathrm{PS}(Y,z)$ in the half-channel $0\leq
    Y \leq 1/\epsilon$, for $\Pran = 300$, contoured with intervals matching
    those in Fig.~\ref{fig:highpr_example}.  The streamlines show the
    streamwise-averaged fluid motion in the $(Y,z)$-plane and are coloured from
    blue (inflow toward the wall), to white ($\bar{v} = 0$), to red (outflow
    from the wall) according to the value of $\bar{v}$.  Regions of outflow
    are shaded grey. 
    (\emph{b})~Asymptotic density profile $\rho_0(Y,z^*)$
    [Eq.~\eqref{eq:rho0int}], (black dotted, computed with $\bar{v}_2 = -36.2$
    to $3$ s.f.), plotted alongside the streamwise-averaged density
    $\bar{\rho}_\mathrm{PS}(Y, z^*)$ at $\Pran = 300$, for the inflow stagnation
    point at $z^*=\pi/2$.
	} 
	\label{fig:rhobar_inout}
\end{figure}

At stagnation points $z^*$ where the spanwise velocity is zero, the right hand
side of Eq.~\eqref{eq:rho0asym} vanishes and it reduces to an ordinary
differential equation for $\rho_0$. A solution of this, which decays as it leaves
the boundary layer ($Y \rightarrow \infty$), is
\begin{equation}%
	\rho_0(Y, z^*) = 1 -  \frac{(9|\bar{v}_2|)^{1/3}}{\Gamma(1/3)}\int^Y_0 \exp\left(\tfrac{1}{3}\bar{v}_2 s^3\right)\,\mathrm{d}s,
	\label{eq:rho0int}%
\end{equation}%
and only exists if $\bar{v}_2 < 0$, i.e.\ if there is inflow to the $y = -1$ wall.
Figure~\ref{fig:rhobar_inout}(\emph{a}) shows a contour plot of
$\bar{\rho}_\mathrm{PS}(Y,z)$ at high Prandtl number, $\Pran = 300$. 
%
%
% Fig 12
%
The density boundary layer is approximately uniform in the spanwise direction
where there is inflow into the boundary layer and the asymptotic solution
[Eq.~\eqref{eq:rho0int}] matches the numerical solution at the inflow stagnation
points: see Fig.~\ref{fig:rhobar_inout}(\emph{b}).

Significantly, there is no equivalent solution of the form in
Eq.~\eqref{eq:rho0int}, for an `outflow' stagnation point. 
This is because these points, $z=\pi/4$ and $3\pi/4$, are the focus of boundary
layer eruptions [see Fig.~\ref{fig:rhobar_inout}(\emph{a})], where the layer
scaling breaks down. These eruptions are the $\Ri_b \to 0$ manifestation of the
fingers seen in Fig.~\ref{fig:bldevel} for $\Ri_b=0.01$. The situation is
clearest at high $\Rey$ where the flow and density field are dominantly
streamwise-independent. Defining a length scale $\delta$ for the spanwise width
of the fingers, the velocity fields may be Taylor expanded about the outflow
stagnation points $(y,z)=(-1,z^*)$ as follows: $\bar{v}(Y,Z)= \epsilon^2
\Rey^{-1} A Y^2 + O(\epsilon^2 \delta, \epsilon^3)$, and (using
incompressibility) $\bar{w}(Y,Z)= -2
\epsilon \delta \Rey^{-1} A Y Z + O(\epsilon^2 \delta, \epsilon \delta^2)$,
where $Z \coloneqq (z-z^*)/\delta$ and $A$ is an $O(\Pran^0)$ constant. Then,
instead of Eq.~\eqref{eq:blstratfull}, we have
\begin{equation} \epsilon A Y^2 \frac{\partial\rho_0}{\partial Y} - 2\epsilon A
    YZ \frac{\partial\rho_0}{\partial Z}  = \frac{1}{\Pran} \left(
    \frac{1}{\epsilon^2} \frac{\partial^2 \rho_0}{\partial Y^2}+
    \frac{1}{\delta^2} \frac{\partial^2 \rho_0}{\partial Z^2} \right).
\end{equation}
%
%Here $Y:=(y+1)/\epsilon$ as before, $Z:=(z-z^*)/\delta$ and the velocity fields
%have been Taylor expanded about the ouflow stagnation point $(y,z)=(-1,z^*)$ as
%follows: $\bar{v}_2(y,z)= \epsilon^2 A Y^2+O(\epsilon^2 \delta, \epsilon^3)$, and
%$\bar{w}_1(y,z)= -2 \epsilon \delta A Y Z + O(\epsilon^2 \delta, \epsilon
%\delta^2)$ ($A$ is an $O(\Pran^0)$ constant).  

As the flow in the boundary layer approaches an outflow stagnation point, it
enters a corner region where the cross-stream diffusion becomes subdominant.
(See \cite{Childress79} and \cite{ChildressGilbert}, pages 135--136, for a
discussion of this for an equivalent magnetic field problem, although note that
there the velocity boundary conditions used there are stress-free rather than
non-slip conditions here.) Therefore, the density field is simply advected by
the flow and the leading balance is
\begin{equation} \epsilon A Y^2 \frac{\partial\rho_0}{\partial Y} - 2\epsilon A
YZ \frac{\partial\rho_0}{\partial Z}  = 0.  \end{equation}
This has a solution of the form $\rho_0=\rho_0(\psi)$ where $\psi \coloneqq
A(\epsilon Y)^2 \delta Z$ is a streamfunction. 

Consider a streamline starting in
the boundary layer where $(\epsilon, \delta)=(\Pran^{-1/3},1)$. Then
$\psi=O(\Pran^{-2/3})$, requiring $\epsilon$ and $\delta$ to satisfy the
condition
\begin{equation} \epsilon^2 \delta = \Pran^{-2/3} \label{condition1}
\end{equation} as the streamline negotiates the corner. In the corner itself,
$\epsilon=\delta$ (this is the scaling of the closest point of approach of
the streamline to the stagnation point), so the corner is defined by the
scalings 
$$\epsilon=\delta=\Pran^{-2/9}.$$

\begin{figure}
	%\input{2dfingers.pgf}
	%\importpgf{.}{2dfingers.pgf}%
	\includegraphics{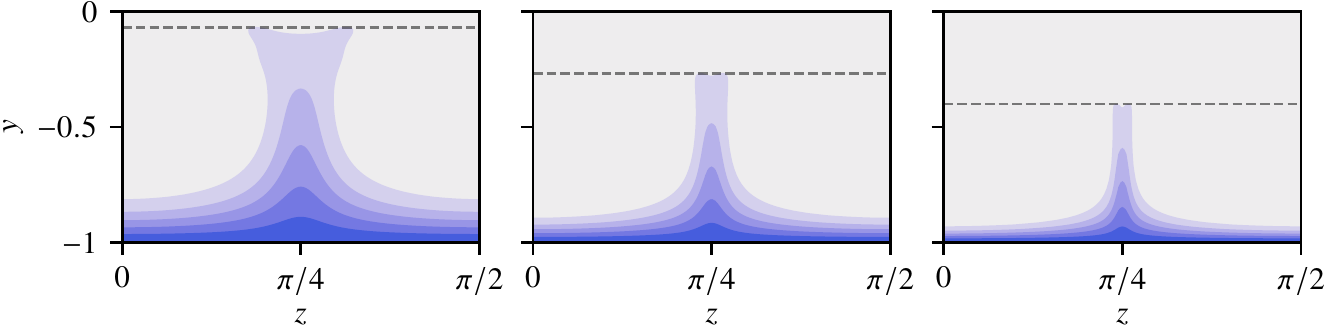}
	\caption{%
        Density fingers diminishing with increasing $\Pran$ 
		in the passive scalar limit.
        Plotted are contours of $\bar{\rho}_\mathrm{PS}(y,z)$ at (from left to
        right) $\Pran = 20$, $100$ and $400$.  The contour intervals match those
        in Fig.~\ref{fig:highpr_example}. We only plot the bottom left quadrant
        of the full channel -- the other parts are dictated by the solution
        symmetries. The dashed grey lines indicate the approximate height of the
        finger structures, at $y=-0.07$, $-0.27$ and $-0.4$ respectively.
	}
	\label{fig:fingershrink}
\end{figure}

\begin{figure}
	%\input{fingerscaling.pgf}
	%\importpgf{.}{fingerscaling.pgf}%
	\includegraphics{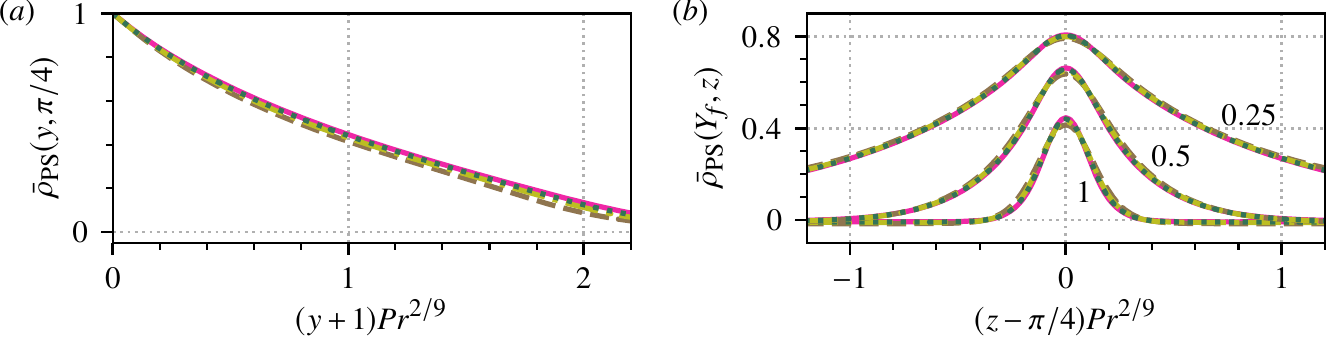}
	\caption{
        Scaling of density fingers at high $\Pran$.  Both parts plot slices
        through the same set of solutions in the passive scalar limit, with
        Prandtl numbers $\Pran = 100$ (brown dashed), $\Pran = 200$ (olive
        dash--dot), $\Pran = 300$ (dark green dotted) and $\Pran = 400$ (magenta
        solid).  
        (\emph{a}) Streamwise-averaged density $\bar{\rho}_\mathrm{PS}$ through
        the finger-like structure at $z = \pi/4$, as a function of the rescaled
        wall-normal co-ordinate $Y_f \coloneqq (y + 1)\Pran^{2/9}$. 
        (\emph{b})~Streamwise-averaged density profiles in the spanwise
        co-ordinate, rescaled in the vicinity of the finger structure at $z =
        \pi/4$ by $\Pran^{2/9}$.  For each Prandtl number, three slices are
        plotted at fixed $Y_f = 0.25$, $0.5$ and $1$ as indicated.  
    }%
    \label{fig:deltascaling}% 
\end{figure}

As the streamline leaves the corner region (with $\delta$ decreasing) to enter
the outer finger region, spanwise diffusion grows to balance  advection, so that
\begin{equation} \epsilon A Y^2 \frac{\partial\rho_0}{\partial Y} -
2\epsilon A YZ \frac{\partial\rho_0}{\partial Z}  = \frac{1}{\Pran
\delta^2} \frac{\partial^2 \rho_0}{\partial Z^2}, \end{equation}
which requires 
\begin{equation} \epsilon \delta^2 = \Pran^{-1}.  \label{condition2}
\end{equation} Combining the conditions \eqref{condition1} and
\eqref{condition2} leads to outer finger scalings of 
\begin{equation}\epsilon= \Pran^{-1/9}, \quad \delta=\Pran^{-4/9}.\label{outerfin} \end{equation}
Significantly, this predicts that the  extent of the fingers' intrusion into the
interior will ultimately vanish (albeit slowly) as $\Pran \rightarrow \infty$ in
this $\Ri_b \to 0$ limit, consistent with the contour plots in
Fig.~\ref{fig:fingershrink}. Unfortunately, the Prandtl numbers reached here are
not large enough to confirm this small exponent. What \emph{is} possible is to
examine the lower finger profile or corner region.
Figure~\ref{fig:deltascaling}(\emph{a}) shows that streamwise-averaged density
profiles through the centreline of the fingers can be collapsed by a rescaled
wall-normal co-ordinate $Y_f \coloneqq (y+1)\Pran^{2/9}$, i.e.\ the lower parts
(corner regions) of the fingers scale like $\Pran^{-2/9}$ in $y$.
Figure~\ref{fig:deltascaling}(\emph{b}) plots spanwise density profiles of
$\bar{\rho}_\mathrm{PS}$, across the finger at $z = \pi/4$, at three fixed $Y_f
= 0.25$, $0.5$ and $1$.  On rescaling the spanwise coordinate by $\Pran^{2/9}$,
the subsequent profiles, ranging from $\Pran = 100$ to $400$, collapse well.
%
% Fig 13
%
%

%--
The important findings from this passive scalar limit are: (\emph{a}) a
rationale for an $O(\Pran^{-1/3})$ density boundary layer being formed;
(\emph{b}) the existence of boundary layer eruptions, producing \emph{fingers},
centred at outflow stagnation points in the boundary layer; and (\emph{c}) the
fact that these eruptions are actually secondary (they diminish with increasing $\Pran$) to the primary  result that the
density gets more and more confined to boundary layers, leaving an homogenised
interior as $\Pran\to \infty$.  We now examine whether this picture holds more
generally for  $\Ri_b >0$ equilibria.

%
% Fig 14
%
%

%-------------------------------------------------------------------------------------------------------
%
%  3.2.2.
%
%-------------------------------------------------------------------------------------------------------
\subsubsection{$\Ri_b > 0$}
\label{sec:increasingri}
\begin{figure}%
    %\input{highprlowri.pgf}
	%\importpgf{.}{highprlowri.pgf}%
	\includegraphics{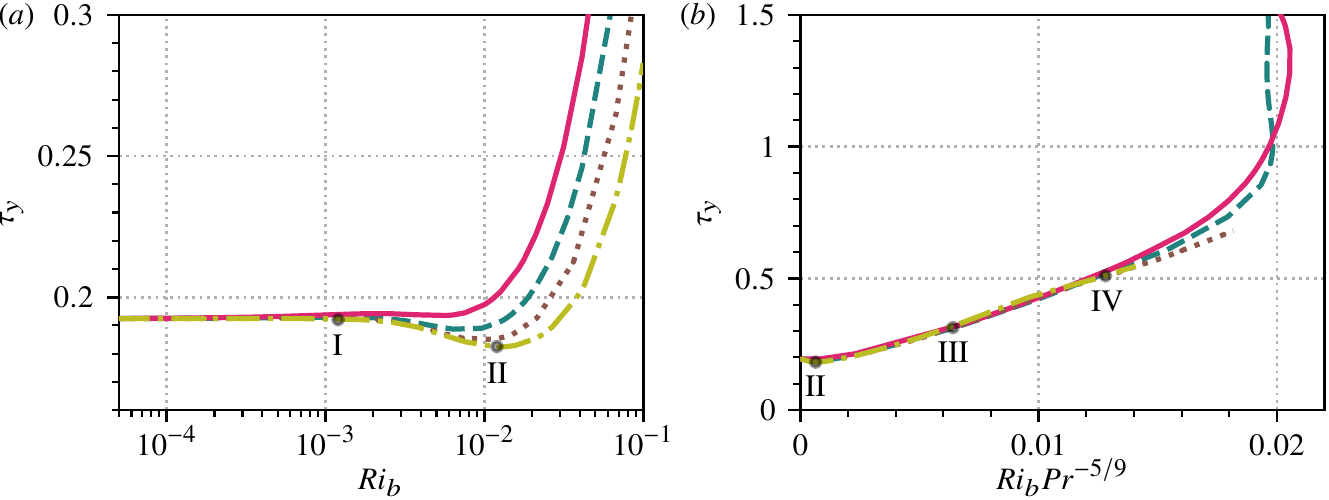}
    \caption{
    Sections of the high-$\Pran$ solution curves from
    Fig.~\ref{fig:curves}(\emph{b}) 
    along the lower branch. Shown are
        $\Pran = 40$ (magenta solid), $70$ (teal dashed), $120$ (brown
        dotted), $200$ (olive dash--dot).
        (\emph{a}) Low $\Ri_b$ portion with logarithmic horizontal axis.
        (\emph{b}) The lower branch plotted with the horizontal axis rescaled by 
        $\Pran^{-5/9}$.
        In both parts, the numerals I--IV indicate the locations of solutions
        plotted below in Figs.~\ref{fig:streaksin}--\ref{fig:lowerbranch_v}.
    }%
	\label{fig:highpr_lowri_curves}%
\end{figure}

We now look at how the lower branch solutions behave as $\Ri_b$ is increased from
zero. In Fig.~\ref{fig:highpr_lowri_curves}(\emph{a}), we replot the lower branch
data from Fig.~\ref{fig:curves}(\emph{b}) on semi-log axes, for low $\Ri_b$ and
$\Pran \geq 40$ (where the boundary layer and finger structures are well
developed). 
%
%
% Fig 15
%
As in both the low $\Pran$ and $\Pran = O(1)$ \citep{Deguchi2017, Olvera2017}
cases, there is necessarily a passive scalar regime where the momentum and
stratification equations [Eqs.~\eqref{eq:momfull} and~\eqref{eq:stratfull}]
remain effectively uncoupled. This is the approximately horizontal part of the
solution branch, from $\Ri_b = 0$ to $\Ri_b \approx 10^{-3}$.  The fact that the
upper limit of this regime does not display any dependence on $\Pran$ contrasts
with the scaling of $\Ri_b=O(\Pran^{-1}\Rey^{-2})$ for $\Pran \lesssim O(1)$
\citep[][and \S\ref{sec:lowpr}]{Deguchi2017, Olvera2017} in which $\Ri_b$ scales
inversely with $\Pran$.  The difference here when $\Pran \gg 1$ is that  the
density perturbation $\hat{\rho}$ has to be $O(1)$ to homogenise the interior.
The buoyancy force is then $O(\Ri_b)$ and the streamwise rolls first feel its
influence when $\Ri_b=O(\Rey^{-2})$.  It is worth noting that this argument
relies on the fact that the interior is only partly homogenised, otherwise
$\hat{\rho}=y$ would precisely cancel out the base state and then the buoyancy
force can be exactly balanced by the pressure.

\begin{figure}%
    %\input{streaksin.pgf}
	%\importpgf{.}{streaksin.pgf}%
	\includegraphics{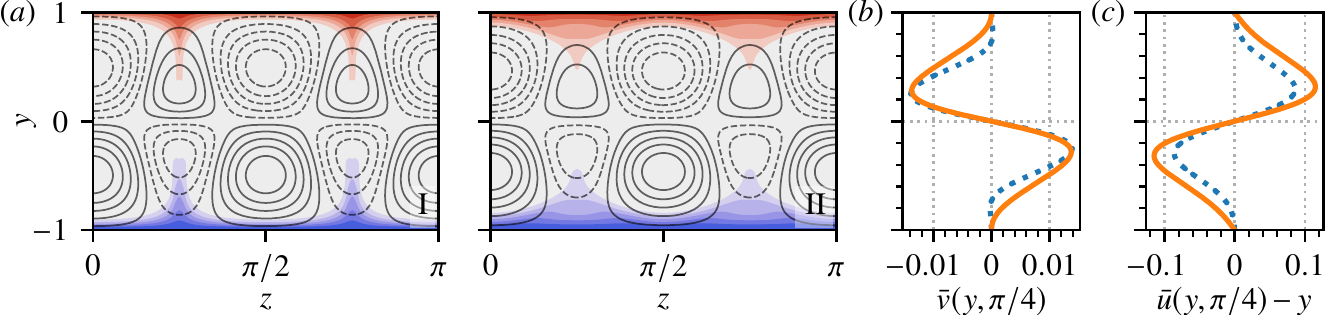}
    \caption{Plots demonstrating the stress drop in the high-$\Pran$ lower
        branch states. 
        (\emph{a})~Comparison of $EQ_7$ at $\Pran = 200$ and $\Ri_b = 1.2\times
        10^{-3}$ (left) and $1.2\times 10^{-2}$ (right), located at points I and II on
        Fig.~\ref{fig:highpr_lowri_curves}(\emph{a}) respectively. The plots show
        $\bar{\rho}$ contours in the $yz$-planes, plotted from red ($\bar{\rho}
        < 0$) to blue ($\bar{\rho} > 0)$,
        using intervals matching those in Fig.~\ref{fig:highpr_example}.
        Overlaid in black lines are 10 isolines of the streamwise-averaged 
        $\hat{u}$ field,
        separated by equally spaced intervals between $\pm 0.2$. Dashed lines
        represent $\bar{u} - y < 0$, solid lines $\bar{u} - y > 0$.
        (\emph{b})~Profiles of $\bar{v}$ for the solutions depicted in part
        (\emph{a}), through $z = \pi / 4$. 
        (\emph{c})~Corresponding profiles of $\bar{u}-y$, through $z = \pi / 4$.
        In parts~(\emph{b}) and~(\emph{c}), solid orange lines are the $\Ri_b =
        1.2\times 10^{-3}$ state (I) and dashed blue lines are the $\Ri_b =
        1.2\times 10^{-2}$ state (II).
    }%
	\label{fig:streaksin}%
\end{figure}

Just beyond $\Ri_b=10^{-3}$ in Fig.~\ref{fig:highpr_lowri_curves}(\emph{a}), the
lower branch solutions all experience a small drop in stress at the walls. The
reason for this is encapsulated in Fig.~\ref{fig:streaksin}(\emph{a}), where
streamwise-averaged density slices through the $yz$-plane are
plotted for the $\Pran = 200$ density field at $\Ri_b = 1.2\times 10^{-3}$
(point I, Fig.~\ref{fig:highpr_lowri_curves}) and $1.2\times 10^{-2}$ (point II,
Fig.~\ref{fig:highpr_lowri_curves}), with streamwise
velocity contours overlaid.
%
% fig 16
%
Between these two bulk Richardson numbers, the four streaks centred at $z =
\pi/4$ and $3\pi/4$ noticeably weaken and recede from the walls. Increasing
$\Ri_b$ penalises the upward motion of dense fluid and likewise the downward
motion of less-dense fluid.  Consequently, we observe in
Fig.~\ref{fig:streaksin}(\emph{b}), that the magnitude of the $\bar{v}$ field
drops significantly for $0.3 \lesssim |y| \leq 1$, in the vicinity of the
density fingers, leading to the weakening of the streaks there. This is
highlighted further in Fig.~\ref{fig:streaksin}(\emph{c}), where we plot
streamwise-averaged slices of $\hat{u}$ through $z = \pi/4$ and see that it
drops along the approximate length of the fingers. The density fingers also
retreat as $\Ri_b$ increases, since they are no longer vertically advected as
strongly. The streaks centred at $z = 0 \equiv \pi$ and $\pi/2$ are largely
unaffected, since they only experience significant buoyancy forces in the thin
boundary layer.  Therefore, there is a small reduction in mean wall stress in
this regime, caused by the streamwise flow receding from the wall in the finger
regions. 

Following the mean stress drop, there is a marked and sustained rise in $\tau_y$
as $\Ri_b$ increases further and states enter a new regime, which covers most of
the lower branch. 
Working on the basis that it is the stratification invading the interior which
exerts the leading influence on the ECS, a simple estimate for how $\Ri_b$
varies with $\Pran$ can be deduced as follows. Assuming the passive scalar
finger scalings, the fingers represent an $O(1)$ stratification perturbation to
the interior in a volume of size $O(1) \times O(\Pran^{-1/9})\times
O(\Pran^{-4/9})$. This is equivalent to an $O(\Pran^{-5/9})$ stratification
over the entire $O(1)$ volume which could be expected to influence the
wall-normal momentum equation when $\Rey^{-2} \sim \Ri_b \Pran^{-5/9}$ (balancing
viscous diffusion with the buoyancy term) or $\Ri_b =O(\Pran^{5/9})$.
%
%By volume averaging along the $O(\Pran^{-1/9})\times O(\Pran^{-4/9})$ extent of
%the density fingers identified in Eq.~(\ref{outerfin}), the spatial extent of the $\rho$ field that impinges most
%significantly on the flow (thereby affecting the wall stress) is
%$O(\Pran^{-5/9})$, suggesting this as a rescaling for the continuation curves.
While the finger scales are by no means guaranteed to persist beyond the
weakly stratified regime, Fig.~\ref{fig:highpr_lowri_curves}(\emph{b})
demonstrates a remarkably good collapse of the data.

\begin{figure}
	%\input{lowerbranch_contours_xbar.pgf}
	%\importpgf{.}{lowerbranch_contours_xbar.pgf}%
	\includegraphics{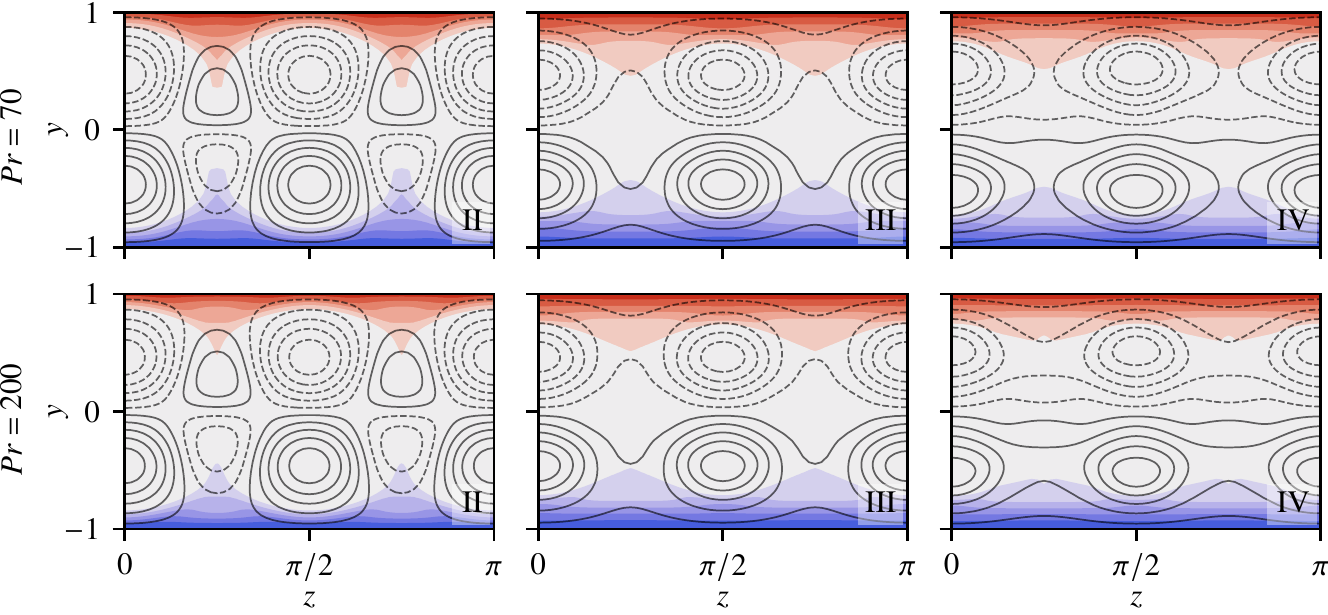}
    \caption{
    Streamwise-averaged contour plots of $EQ_7$ at high $\Pran$ along the 
    ($\Pran^{-5/9}$-rescaled) lower
    branch, at the points II--IV, as indicated in
    Fig~\ref{fig:highpr_lowri_curves}(\emph{b});
    %Streamwise-normal contour plots for $EQ_7$ through $x=0$ at high $\Pran$,
    %increasing in Richardson number, along the lower branch, rescaled by
    %$\Pran^{-1/2}$. 
    $\Ri_b
    \Pran^{-5/9} = 6.4\times 10^{-4}$ (II), $6.4\times 10^{-3}$ (III)
    and $1.28 \times 10^{-2}$ (IV), with $\Pran = 70$ (top row) and $200$ (bottom
    row).
    Contours of $\bar{\rho}(y,z)$ are shown, plotted from red ($\bar{\rho} > 0$)
    to blue ($\bar{\rho} < 0$); intervals match those in
    Fig.~\ref{fig:highpr_example}. Overlaid in black are 10 evenly spaced
    isolines of the streamwise average of $\hat{u}$, between $\pm 0.2$ (II),
    $\pm 0.29$ (III) and $\pm 0.37$ (IV). Dashed lines are negative contours and
    solid lines are positive.
    For reference, the bulk Richardson numbers for the $\Pran = 70$ data are
    $6.8\times 10^{-3}$ (II), $6.8\times 10^{-2}$ (III), and $0.14$ (IV).
    For the $\Pran = 200$ data, $\Ri_b = 1.2\times 10^{-2}$ (II), $0.12$ (III),
    and $0.24$ (IV).
    }
	\label{fig:lowerbranch_contours_xbar}
\end{figure}

Figure~\ref{fig:lowerbranch_contours_xbar} shows $yz$-contours of the
streamwise-averaged streak field $\bar{u}-y$, overlain on $\bar{\rho}$, for
states at three stages along the lower branch, namely the (rescaled) locations
labelled on Fig.~\ref{fig:highpr_lowri_curves}(\emph{b}): $\Ri_b \Pran^{-5/9} =
6.4 \times 10^{-4}$ (point II, at the stress drop), $6.4 \times 10^{-3}$ (point
III) and $1.28\times 10^{-2}$ (point IV).
%
% Fig 17
%
The similarity between the $\Pran=70$ and the $\Pran=200$ fields  is immediately
striking, with the principal difference only being that the $\Pran = 200$
density field is slightly more homogenised in the interior. In both cases, we
see the streak field developing in a similar way to the low $\Pran$ states (see
Fig.~\ref{fig:lowpr}); the wall outflow streaks at $z=\pi/4$ and $3\pi/4$ recede
into the interior, become dominated by the inflow streaks at $z = 0 \equiv \pi$
and $\pi / 2$ (point III), and the streamwise velocity perturbation separates
into negative and positive halves (point IV).

\begin{figure}
	%\input{lowerbranch_v_xbar.pgf}
	%\importpgf{.}{lowerbranch_v_xbar.pgf}%
	\includegraphics{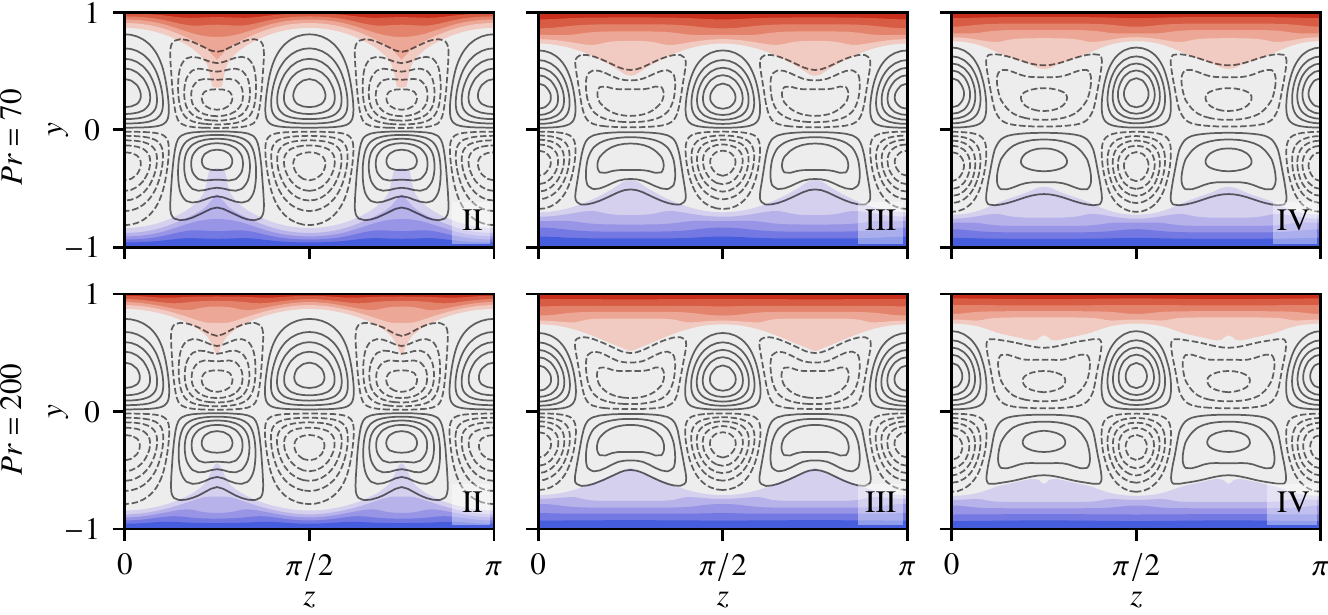}
    \caption{
    Evolution of the $\bar{v}$ field for increasing $\Ri_b$ along the
    high-$\Pran$ lower branches, $\Pran= 70$ (top row) and 200 (bottom row).
    Each pane features $10$ equally spaced isolines of $\bar{v}$ corresponding
    to the states shown in Fig.~\ref{fig:lowerbranch_contours_xbar}, located at
    points II--IV along the solution branches in
    Fig.~\ref{fig:highpr_lowri_curves}(\emph{b}). The isolines lie between $\pm
    0.015$ (II), $\pm 0.031$ (III) and $\pm 0.05$ (IV), with solid lines
    indicating $\bar{v} > 0$ and dashed lines $\bar{v} < 0$.
%    Contours of $\hat{v}(0, y, z)$ for the $\Pran = 70$ solutions in
%    part (\emph{a}); 10 evenly spaced contours between (left to right) $\pm
%    1.8\times 10^{-2}$, $\pm 6.1\times 10^{-2}$ and $\pm 0.12$ are used. The
%    background is shaded according to the value of $\hat{v}$, from $-0.12$
%    (pink) to $0$ (white) to $0.12$ (cyan).
    Underneath, the $\bar{\rho}(y,z)$ contours (shown in
    Fig.~\ref{fig:lowerbranch_contours_xbar}) are replotted for reference.
    The bulk Richardson numbers for the $\Pran = 70$ data are $6.8\times
    10^{-3}$ (II), $6.8\times 10^{-2}$ (III), $0.14$ (IV) and for the $\Pran =
    200$ data: $\Ri_b = 1.2\times 10^{-2}$ (II), $0.12$ (III) and $0.24$
    (IV). 
    }
    \label{fig:lowerbranch_v}
\end{figure}

The $\bar{v}$ fields corresponding to the
Fig.~\ref{fig:lowerbranch_contours_xbar} lower branch states are plotted in
Fig.~\ref{fig:lowerbranch_v}, again with $\bar{\rho}$ underlaid.
At the stress drop (II), the $\bar{v}$ field covers the full interior, though it
is somewhat inhibited in the finger regions. By point III (one third along the
lower branch) it has pulled away completely from the highly stratified regions
at the walls and remains that way as $\Ri_b$ increases to point IV (two thirds
along the lower branch).
Meanwhile, the density fingers have smeared out and the stratified layer has
thickened, due to readjustment of the advective-diffusive balances there. It is
worth noting that the magnitude of $\bar{v}$ in the interior has increased
substantially.  This leads to the growth of the streak fields and the corresponding
mean stress increases along the lower branch (see
Figs.~\ref{fig:highpr_lowri_curves}(\emph{b}) and
\ref{fig:lowerbranch_contours_xbar}). However, the main message is that
faced with increasing stratification at the walls, the $\bar{v}$ field isolates
itself, pulling in towards the homogenised interior, where $\bar{\rho}\approx 0$.
This appears to be a key mechanism via which the ECS extends the range of
$\Ri_b$ at which it can persist before being disrupted.  As $\Pran$ increases for any
fixed $\Ri_b$, the stratified layer recedes further into the walls and the
effect of $\Ri_b$ on the structure is less severe. 
Indeed, our data and associated scaling argument hints that the maximum bulk
Richardson number attained by states is $\Ri_b^m = O(\Pran^{5/9})$, though it is
currently challenging to verify this concretely, given the numerical
difficulties inherent in resolving high-$\Pran$ states and their shrinking
length scales.
%indicate that $\Ri_b \Pran^{-5/9} \sim O(1)$ is the scale at which
%states can no longer tolerate increasing global stratification.

%---------------------------------
%
% 3.2.3
%
%---------------------------------
\subsubsection{Other solutions}
\label{sec:othersolns}
While~\cite{Olvera2017} isolated $EQ_7$  \& $EQ_8$ by tracking the edge
manifold, many other equilibria exist in plane Couette flow that are dynamically
important in the unstratified setting.  Since these states are based upon the
same SSP/VWI tripartite structure of rolls, streaks and waves, the expectation
is that our findings of density homogenisation in the interior with stably
stratified boundary layers forming as  a consequence at high Prandtl number are
generic. To confirm this, we continued solutions $EQ_1$--$EQ_{11}$ from
\citet{Gibson2009} to increasing $\Pran$ in the passive scalar limit ($\Ri_b
\rightarrow 0$) and observed how they change as density transport becomes
increasingly convectively dominated.
%
% Fig 19
%

\begin{figure}
	%\input{othersolns.pgf}
	%\importpgf{.}{othersolns.pgf}%
	\includegraphics{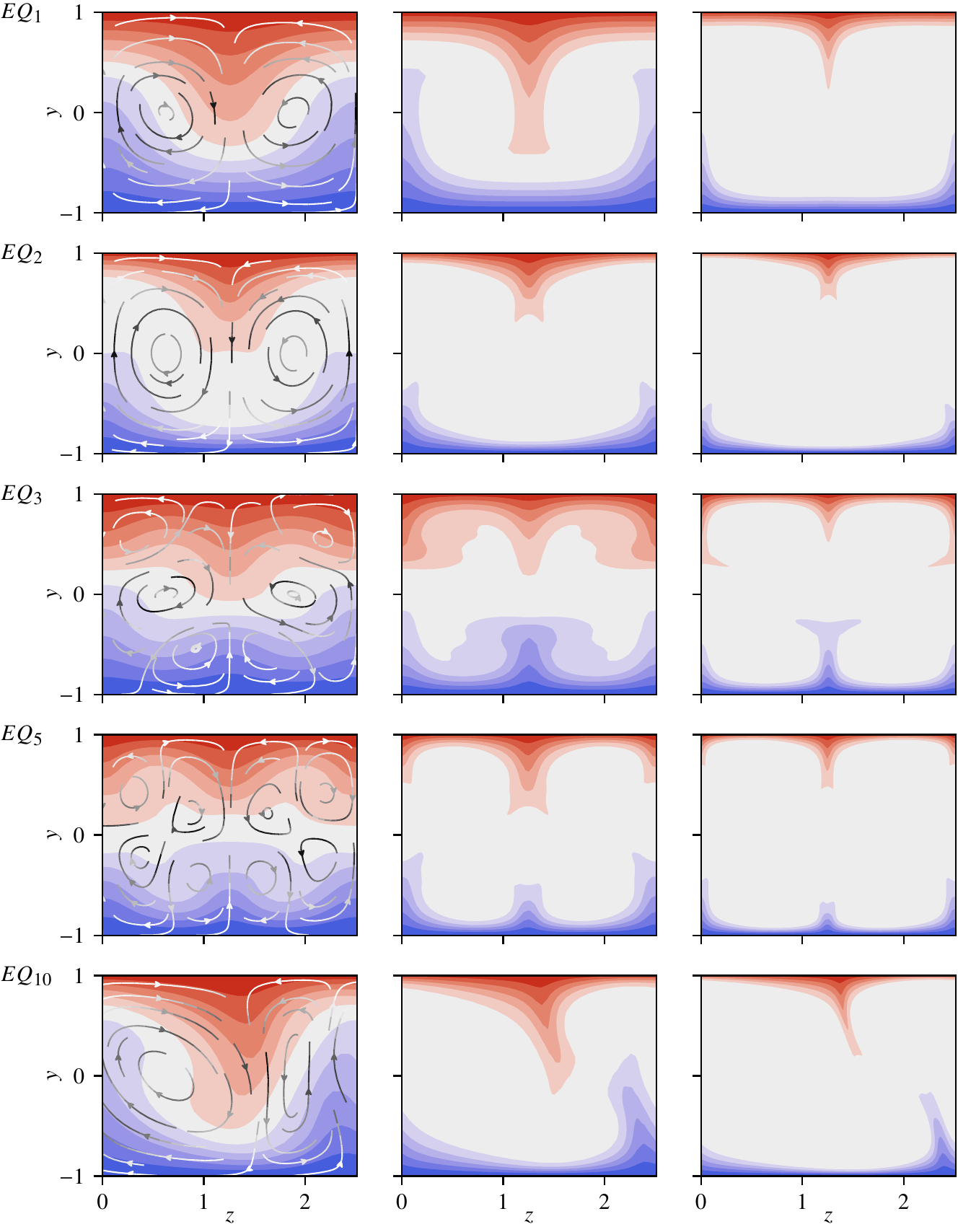}
	\caption{Streamwise-averaged density fields $\bar{\rho}_\mathrm{PS}(y,z)$
for various equilibria in the passive scalar limit.  The columns increase in
Prandtl number from left to right: $\Pran = 1$, $10$ and $70$.  The contour
intervals match those in Fig.~\ref{fig:highpr_example}.  Streamlines of
$(\bar{v},\bar{w})$ are shown on the leftmost plots, coloured with a linear
gradient from white to black, according to $|(\bar{v}, \bar{w})|$, which has a
maximum value of (to $2$ significant figures): $0.018$ ($EQ_1$), $0.098$
($EQ_2$), $0.033$ ($EQ_3$), $0.053$ ($EQ_5$) and $0.061$ ($EQ_{10}$).
	}
	\label{fig:othersolns}
\end{figure}
Figure~\ref{fig:othersolns} shows contours of $\bar{\rho}_\mathrm{PS}(y,z)$ for
a subset of our converged solutions at $\Pran = 1$ (lefthand column), $10$
(middle column) and $70$ (righthand column), with overlaid streamlines on the
lefthand column depicting the different roll structures. Just as for $EQ_7$ in
\S\ref{sec:ri0}, each of these new solutions develops a homogenised interior
with a corresponding highly stratified boundary layer at the walls. Importantly,
each solution features characteristic density fingers, advected into the interior by the
rolls, which shrink as $\Pran$ increases to leave an increasingly homogenised
interior.  The solutions $EQ_4$, $EQ_6$, $EQ_9$ and $EQ_{11}$ (not shown) behave
similarly.

%
%
%   Discussion
%
%
\section{Discussion}
\label{sec:discussion}
%
% What we have done - summary findings
%
In this study, we have computed exact coherent structures (ECS) in stratified
plane Couette flow across a wide range of Prandtl numbers at a fixed Reynolds
number $\Rey=400$.  In the two asymptotic limits $\Pran \ll 1$ and $\Pran \gg
1$, states persist to arbitrarily high bulk Richardson numbers, in spite of the
stabilising influence of stratification. However, the underlying mechanisms
which allow this in each case are quite different. 

In the $\Pran \rightarrow 0$ limit, density transport is diffusion dominated
with perturbations from the constant base density gradient being only
$O(\Pran)$. To leading order, the bulk Richardson number $\Ri_b$ only affects
the ECS in the combination $\Ri_b \Pran$ with stratification having no
noticeable effect on the flow until $\Ri_b =O(\Pran^{-1} \Rey^{-2})$. Beyond
this regime, states continue to follow an $O(\Pran)$ asymptotic solution curve,
reaching a maximum value of $\Ri_b^m \approx 0.01/Pr$ at $\Rey=400$, for the ECS
studied in detail here ($EQ_7$).

In the $\Pran \rightarrow \infty$ limit, density transport is advection
dominated for any $\Ri_b>0$, with the interior becoming homogenised as the
stratification is confined to boundary layers at the walls (their thickness
being $O(Pr^{-1/3})$ in the $\Ri_b \rightarrow 0$ limit).  The flow and the
density stratification are then separated into different parts of the domain
with the stratification only starting to influence the velocity field when
$\Ri_b=O(\Rey^{-2})$ independently of $\Pran$, which differs from the $\Pran
\lesssim O(1)$ scaling of $O(\Pran^{-1} \Rey^{-2})$
\citep{Deguchi2017,Olvera2017}. The maximum stratification which can be
tolerated by the ECS looks to be  $\Ri_b^m \sim O(\Pran^{5/9})$ at fixed $\Rey$.
Numerically, the situation at large but finite $\Pran$ and small $\Ri_b$ is
complicated by eruptions in the density boundary layers at regions of outflow.
These eruptions give rise to \emph{fingers}, which ultimately vanish to leave a
homogenised interior in the limit $\Pran \rightarrow \infty$. This behaviour is
qualitatively captured in the passive scalar limit ($\Ri_b \rightarrow 0$) where
these fingers have spanwise width $O(\Pran^{-2/9})$ near the wall, contracting
to $O(\Pran^{-4/9})$ further away and wall-normal length $O(\Pran^{-1/9})$.

% 
%  Implications
% 

As discussed in the introduction, a dynamical systems perspective of stratified
turbulence imagines a `turbulent' trajectory guided through a
complicated phase space 
%populated 
by a hierarchy of simple invariant solutions
(ECS) 
and their entangled stable and unstable manifolds.
%interconnected through a tangle of all their stable and unstable manifolds
Therefore, it is reasonable to expect that understanding the structure of ECS as
a function of the parameters $\Rey$, $\Pran$ and $\Ri_b$ may help explain what is
seen (maybe fleetingly) in stratified turbulence. Here, we have found simple
(laminar) realisations of boundary layers and homogenised regions seen in
turbulence with clear scalings emerging, at least for small $\Ri_b$. For example,
the $O(\Pran^{-1/3})$ boundary layers found here resonate with empirical
observations \citep{Kader1981,Schlichting2016} and recent numerical work
\citep{Zhou2017} concerning the thickness of the conductive sublayer. Further
work is obviously needed to push these to higher $\Ri_b$ and $\Rey$, but at least
this is a start. 

Perhaps most noteworthy is the homogenisation seen at high $\Pran$, where the
shear-driven flow simply clears the stratification out of its way. This limit is
costly to simulate at high $\Rey$ since the P\'{e}clet number ($\Pen \coloneqq
\Rey\Pran$) is even larger and hence is little explored. However, hopefully our
findings here will help encourage more effort to reach this
environmentally-relevant limit, since these results clearly highlight the
difference between $\Pran \sim 1$ and $\Pran$ being large, where a value of $70$
(let alone $700$ for salt in water) is significant enough to see the difference.

%\JL{I've not put anything in yet but think we should mention that passive
%scalars are interesting in their own right. And much more accessible in DNS}

%One way to test this in our stratified solutions this is at the boundary layer.
%In stratified turbulent shear flows at high $\Pran$, the thickness of the
%conductive sublayer $\delta_c$ is difficult to determine, since experiments must
%be able to effectively characterise turbulent intensity over a vanishingly small
%length scale at the wall~\citep{Davidson2015}.
%
%Consequently, many empirical scaling laws ranging from $\delta_c =
%O(\Pran^{-1/4})$ \citep{Townsend1980} to $O(\Pran^{-1/2})$
%\citep{Na1999,Na2000} and various values in between
%\citep{Shaw1977,Kader1981,Schwertfirm2007,Schlichting2016} have been proposed
%over the years \JL{double check all these}, derived either from experimental
%data or DNS, of passive scalar transport or fully stratified flow.
%
%In our analysis, the diffusive region of equilibria in the Boussinesq
%equations is necessarily confined to generic $O(\Pran^{-1/3})$ layers at the
%wall as $\Pran\to\infty$, thereby supporting proponents of $\delta_c =
%O(\Pran^{-1/3})$ 

%
% The Future
%

\begin{figure}
    %\input{highre.pgf}
	%\importpgf{.}{highre.pgf}%
	\includegraphics{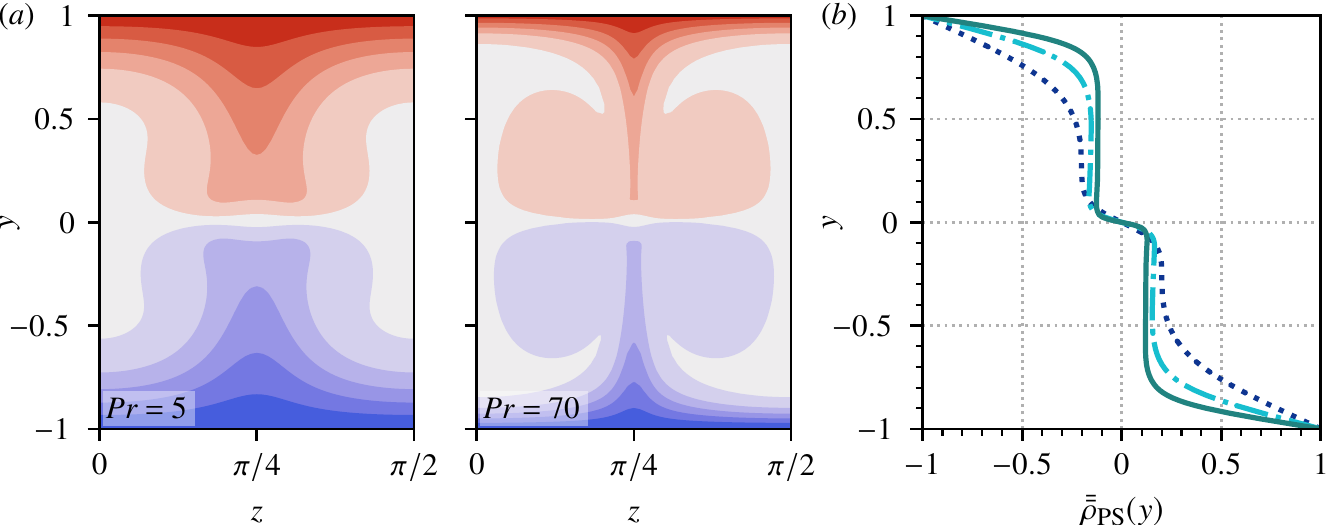}
    \caption{
    Emergence of internal density interfaces as $\Pran$ increases at $\Rey = 10^5$.
    (\emph{a})~Streamwise-normal contour plots of $\bar{\rho}_\mathrm{PS}(y,
    z)$, for $\Pran = 5$ (left) and $\Pran = 70$ (right). Half the channel ($0
    \leq z \leq \pi/2$) is shown (the rest being dictated by symmetry).
    (\emph{b})~Spanwise- and streamwise-averaged density profiles
    $\bar{\bar{\rho}}_\mathrm{PS} \coloneqq
    \frac{1}{L_z}\int_0^{L_z}\bar{\rho}_\mathrm{PS}(y,z)\,\mathrm{d}z$, for
    $\Pran = 5$ (blue dotted), $20$ (turquoise dash-dot) and $70$ (teal solid).
    }
	\label{fig:highre}
\end{figure}

Looking ahead, the existence of highly stratified boundary layers found here in
the large $\Pran$ limit begs the question whether stably-stratified
\emph{interior} density interfaces could exist for these unstable steady flows.
One exploratory computation, taking the $EQ_7$ solution in the passive scalar
limit and increasing $\Rey$ from 400 to $10^5$ indicates `yes'.
Fig.~\ref{fig:highre}(\emph{a}) plots contours of the resulting
streamwise-averaged density fields and, from $\Pran = 5$ to $70$, we see the
usual finger structures and boundary layers developing. In this case,
however, the top and bottom halves of the channel interior have separated into
distinct patches of roughly constant density. In this high-$\Rey$ regime, the
ECS converges to a VWI state, becoming streamwise-invariant to leading order
(see \S\ref{sec:ecstheory}).  The symmetries of $EQ_7$ dictate that $v(x,0,z)\to
0$ as $\Rey \to \infty$ (since the flow becomes increasingly streamwise
independent and $v(x,0,z)=-v(-x,0,z)$).  Therefore, while density in the top and
bottom half-channels can become well-mixed as $\Pran$ increases, there is little
exchange between the two. (Note that states such as $EQ_1$, whose rolls span the
full height of the channel, do not enjoy this property).

\begin{figure}
    %\input{loglog.pgf}
	%\importpgf{.}{loglog.pgf}%
	\includegraphics{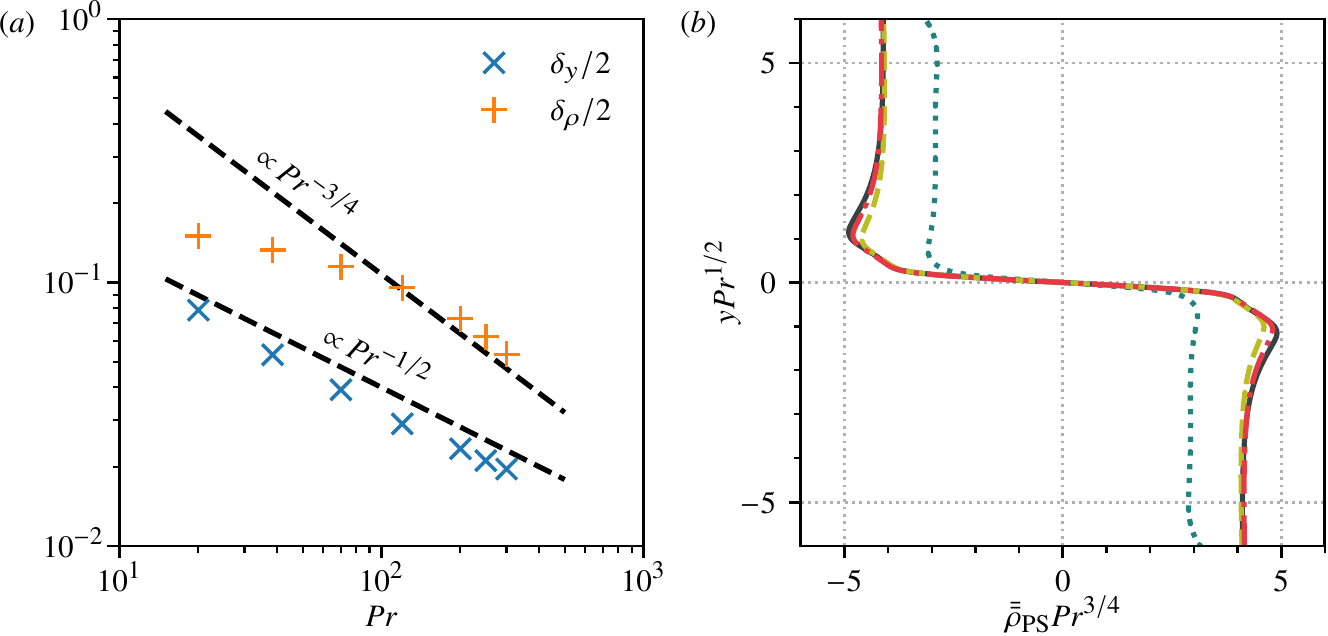}
    \caption{%
    Empirical scalings of the high-$\Rey$ ($=10^5$), high-$\Pran$ internal
    layers plotted in Fig.~\ref{fig:highre}.
    (\emph{a})~Log-log plot showing half the interface thickness $\delta_y$
    (blue~$\times$) and half the density change $\delta_\rho$ (orange~$+$)
    between the two well-mixed regions, as a function of $\Pran$, ranging from
    $20$ to $300$. The two black dashed lines are $0.4\Pran^{-1/2}$ and
    $3.4\Pran^{-3/4}$. Across all data points, the interface is contained within
    the region $\mathcal{I}\coloneqq (-0.2,0.2)$ and $\rho_\mathrm{PS}$ reaches
    an approximately constant value near the edges of this region.  Therefore,
    the half-thickness $\delta_y/2$ was estimated by
    $\operatorname{arg\,max}_{y\in\mathcal{I}} \bar{\rho}_\mathrm{PS}(y,
    3\pi/8)$ and $\delta_\rho/2$ was estimated using
    $\bar{\rho}_\mathrm{PS}(-0.2, 3\pi/8)$.
    (\emph{b})~Streamwise- and spanwise-averaged density profiles
    $\bar{\bar{\rho}}_\mathrm{PS}(y)$, rescaled as indicated, by the
    observed $\Pran$-dependences in part (\emph{a}).  The Prandtl numbers are
    $\Pran = 70$ (teal dotted), $200$ (olive dashed), $250$ (red dash-dot) and
    $300$ (dark grey solid).
    }
    \label{fig:loglog}
\end{figure}

Fig.~\ref{fig:highre}(\emph{b})  plots the spanwise average of
$\bar{\rho}_\mathrm{PS}(y,z)$ against $y$ for $EQ_7$ and reveals a highly
stratified region at the midplane. In this passive scalar limit where the
velocity field is independent of $\Pran$, an advective-diffusive balance sets
the thickness of this interface at $O(\Pran^{-1/2})$ which is borne out  by the
limited data here, plotted in Fig.~\ref{fig:loglog}(\emph{a}) up to $\Pran =
300$. The scaling of the density change across the layer is less clear, though
$O(\Pran^{-3/4})$ seems to capture its dependence for $\Pran \gtrsim 100$ and
taken together, these two scales succeed in collapsing the density profiles at
the interface in Fig.~\ref{fig:loglog}(\emph{b}). Whilst this ultimately implies
that the interface vanishes in the $\Pran \to \infty$ limit, it is nevertheless
finite in the $\Pran = O(10^2)$ range most relevant to geophysical phenomena.
Clearly more computations need to be done to explore this phenomenon. 

\acknowledgements{
The numerical component of this study was carried out using the computational
facilities of the Advanced Computing Research Centre, University of Bristol
(http://www.bris.ac.uk/acrc/).
JL acknowledges the support of an EPSRC Doctoral Prize fellowship under grant
number EP/N509619/1 and the Channelflow 2.0 developers, for assistance with
their code.
RRK acknowledges the support of EPSRC under grant
number EP/K034529/1. 

}

%
%     A
%
\appendix%
\section{Numerical code details}\label{appendix:numerics}%
The flow fields in our direct steady Boussinesq solver are represented by
the real part of the truncated pseudo-spectral expansion
%Any $L_x$- and $L_z$-periodic field $\xi : [0, L_x] \times [-1,1] \times [0, L_z] \to
%\mathbb{R}$ may be represented by the real part of
%
\begin{equation}
    \sum_{m = -M}^M \sum_{n = 1}^N \sum_{l = -L}^L
    \xi_{n,l,m} F_n(y)
   % \xi(x,y,z) = 
%    \sum_{m=-\infty}^\infty\sum_{n=0}^\infty
%    \sum_{l=-\infty}^\infty \xi_{nlm} T_n(y)
	\mathrm{e}^{i(m\alpha x + l\beta z)}
    \label{eq:generalfield}
\end{equation}
for $(x,y,z)\in [0,L_x]\times[-1,1]\times[0,L_z]$, where $\alpha = 2\pi / L_x$ and
$\beta = \pi / L_z$. The $\xi_{n,l,m}$ are unknown complex coefficients and $F_n
\coloneqq F_n^{(r)}+iF_n^{(i)}$ are basis functions in the wall-normal
co-ordinate chosen to satisfy the boundary conditions.  For $\xi \in \{\hat{u},
\hat{v}, \hat{w}, \hat{\rho}\}$, we have $\xi(\pm 1) = 0$ and use
\begin{equation}
    F_n^{(r)} = T_{2n+1} - T_{2n-1}, F_n^{(i)} = T_{2n} - T_{2n-2},
    \label{eq:phibasis}%
\end{equation}
where $T_n$ denotes the $n$-th Chebyshev polynomial; for the pressure field we
use
\begin{equation}
    F_n^{(r)} = T_{2n-1}, F_n^{(i)} = T_{2n-2}.
    \label{eq:psibasis}%
\end{equation}
%
%, $\xi_{nlm}$ are complex
%coefficients, $\alpha \coloneqq 2\pi / L_x$ and $\beta \coloneqq 2\pi / L_z$.
%, since the Cartesian product of the Chebyshev polynomials and
%Fourier modes in $x$ and $z$ is a basis for the space. 
%To discretise such fields in a pseudo-spectral code, the sums in
%Eq.~\eqref{eq:generalfield} are truncated up to maximum wavenumbers $M$, $L$ in
%the $x$ and $z$ (Fourier) directions respectively and a maximum polynomial
%degree $N$ in the $y$ (Chebyshev) direction. 
%
We save computer memory by omitting coefficients that are fixed by the
symmetries defined in Eqs.~\eqref{eq:shiftrefl}--\eqref{eq:spanrefl}.
%
%This is achieved by modifying the representation in Eq.~\eqref{eq:generalfield}.
%In particular, the $\spanrefl$ symmetry is obtained by replacing
%$\mathrm{e}^{il\beta z}$ with $\cos(l\beta z)$ in the $u$, $v$, $p$ and $\rho$
%fields and with $\sin(l\beta z)$ in the $w$ field.  \JL{$\zrot$ and the BCs are
%obtained by replacing $T_n$ with bespoke basis} \JL{$\shiftrefl$ is obtained by
%omitting all modes for which $m+l$ is odd.}
In particular, $\shiftrefl$ implies that $\xi_{n,l,m} = 0$ if $m + l$ is odd;
$\spanrefl$ implies that $\xi_{n,l,m} = \xi_{n,-l,m}$ for $\xi \in \{u, v, p,
\rho\}$ and $\xi_{n,l,m} = -\xi_{n,-l,m}$ for $\xi = w$; $\zrot$, together with
our basis choices in Eqs.~\eqref{eq:phibasis} and~\eqref{eq:psibasis} leads to
%, we
%replace the usual Chebyshev polynomials with complex basis functions $F_n =
%F_n^o + iF_n^e$, where $F_n^o$ and $F_n^e$ are respectively odd and even
%functions of $y$. This then implies 
%that 
$\xi_{n,l,m}^{} = \xi^*_{n,l,-m}$ for $\xi \in \{u, v, p, \rho\}$ and
$\xi_{n,l,m}^{} = -\xi^{*}_{n,l,-m}$ for $\xi = w$, where the asterisk denotes complex
conjugation. These relations reduce the number of Fourier modes required from
$(2M+1)(2L+1)\approx 4ML$ to $\left\lfloor\frac{1}{2}((M+1)(L+1)+1)\right\rfloor \approx ML / 2$.

%Equilibria are converged by collocating the steady form of
Equilibria are converged by inverting the steady form of
%Eqs.~\eqref{eq:momfull}--\eqref{eq:stratfull} over the Chebyshev points $y_j$
Eqs.~\eqref{eq:momfull}--\eqref{eq:stratfull} for each Fourier mode at the
Chebyshev collocation points $y_j$, defined by
\begin{equation}
    -1 < y_j \coloneqq \cos \left[ \frac{\pi (2(j+N) -1)}{4N} \right] < 0,
\quad j = 1, \ldots, N.
\end{equation}
%
%and inverting the resulting system of nonlinear equations.  
Note that the
remaining half-channel, $0 \leq y \leq 1$, need not be discretised, since it is
given by $\zrot$.

Solution branches were numerically refined until increasing $M$, $N$ or $L$ no
longer noticeably affected $\tau_y$. Individual states plotted for analysis were
typically refined further to ensure that the density field was well-resolved.
Typical resolutions were $(M,N,L) = (12,40,12)$ in the low-$\Pran$
case (see \S\ref{sec:lowpr}) and up to a maximum of $(12,90,38)$ or $(10,120,30)$
for high-$\Pran$ states analysed in \S\ref{sec:highpr}. In the passive scalar limit
$\Ri_b\to 0$, the velocity and pressure fields are independent of density.
Consequently, we need only solve Eq.~\eqref{eq:stratfull} subject to fixed $u$,
$v$ and $w$ fields. This method was to verify the structure scalings derived in
\S\ref{sec:ri0} and to compute the high-$\Rey$ states discussed in
\S\ref{sec:discussion}, achieving resolutions for $\rho$ of up to $(18,140,60)$
and $(4,650,68)$ respectively.
In all cases the system possessed roughly $200\,000$ degrees of freedom at
maximum resolutions, requiring $\approx 400$Gb of computer memory to directly
invert the linear operator.

Details of the Channelflow code used to compute states in \S\ref{sec:othersolns}
may be found elsewhere \citep{Gibson2008,Channelflow}. This base code was
modified to include Eq.~\eqref{eq:stratfull} in the time stepping routines. The
maximum resolution employed was to converge $EQ_3$ at $\Pran = 70$ and used a
discretised box with $(N_x,N_y,N_z) = (120,221,189)$ physical points in the $x$,
$y$ and $z$ directions. After omitting dealiased modes this corresponds to
maximum Fourier wavenumbers of $39$ in $x$, $61$ in $z$ and $110$ collocation
points in the half-channel $y\in(-1,0)$.

%
%     B
%
\section{Streamwise-invariance of the high-$\Pran$ boundary layer}\label{appendix:bl}
A steady passive scalar layer of thickness $\epsilon$ at the wall in
(incompressible) plane Couette flow must be streamwise-invariant to leading
order. 
To show that this is true we assume the opposite and reach a contradiction.
Substituting the asymptotic expansions of
Eqs.~\eqref{eq:uexpan}--\eqref{eq:wexpan} into the steady form of
Eq.~\eqref{eq:stratfull} and presuming $\partial_x \rho_\mathrm{PS} =
O(\epsilon^0)$ leads to the dominant balance
\begin{equation}
    -\frac{\partial \rho_\mathrm{PS}}{\partial x} 
	= 
    \frac{1}{\epsilon^2\Rey\Pran} \frac{\partial^2\rho_\mathrm{PS}}{\partial Y^2}.
	\label{eq:wrongleading}
\end{equation}
Taking a Fourier transform of Eq.~\eqref{eq:wrongleading} in $x$ yields
$-ik_x\epsilon^2\Rey\Pran\tilde{\rho}_\mathrm{PS} =
\partial_{YY}\tilde{\rho}_\mathrm{PS}$ for a given streamwise wavenumber $k_x$,
where $\tilde{\rho}_\mathrm{PS}(k_x, Y, z,t)$ is the transformed density field.
At any particular $z$, this has the general solution
\begin{equation}
    \tilde{\rho}_\mathrm{PS}(k_x; Y,z,t) 
	= 
	C_1 \mathrm{e}^{Y\sqrt{-ik_x\eta}} + C_2 \mathrm{e}^{-Y\sqrt{-ik_x\eta}}
	\label{eq:wrongrhohat}
\end{equation}
for constants $C_1$ and $C_2$ and $\eta := \epsilon^2\Rey\Pran$. For the density
field not to blow up in the interior, $C_1=0$, leaving just $C_2$ to be
determined.  Since the density boundary condition at the wall has no streamwise
variation, $\tilde\rho_\mathrm{PS}(k_x; 0,z,t) = 0$ for any $k_x \neq 0$.
Consequently, $C_2$ also has to vanish meaning that $\rho_\mathrm{PS}$ has no
streamwise variation, which is a contradiction.

\bibliographystyle{jfm}
% Note the spaces between the initials
\bibliography{ECS}

\end{document}